\documentclass[aps,pre,reprint,superscriptaddress]{revtex4-1}

\pdfoutput=1

\usepackage{graphicx}
\usepackage{amssymb}
\usepackage{amsmath}
\usepackage{xcolor}
\usepackage{pdfpages}
\usepackage{pgffor}
\usepackage{CJK}
\usepackage{multirow}
\usepackage{gensymb}

\makeatletter
\AtBeginDocument{\let\LS@rot\@undefined}
\makeatother

\begin{document}

\title{Hydration, Ion Distribution, and Ionic Network Formation in Sulfonated Poly(arylene ether sulfones)} %Title of paper

\author{Britannia Vondrasek}
\affiliation{Macromolecules Innovation Institute, Virginia Polytechnic Institute and State University, Blacksburg, Virginia 24061, USA}
\author{Chengyuan Wen}
\affiliation{Macromolecules Innovation Institute, Virginia Polytechnic Institute and State University, Blacksburg, Virginia 24061, USA}
\affiliation{Department of Physics and Center for Soft Matter and Biological Physics, Virginia Polytechnic Institute and State University, Blacksburg, Virginia 24061, USA}
\author{Shengfeng Cheng}
\email{chengsf@vt.edu}
\affiliation{Macromolecules Innovation Institute, Virginia Polytechnic Institute and State University, Blacksburg, Virginia 24061, USA}
\affiliation{Department of Physics and Center for Soft Matter and Biological Physics, Virginia Polytechnic Institute and State University, Blacksburg, Virginia 24061, USA}
\affiliation{Department of Mechanical Engineering, Virginia Polytechnic Institute and State University, Blacksburg, Virginia 24061, USA}
\author{Judy S. Riffle}
\affiliation{Macromolecules Innovation Institute, Virginia Polytechnic Institute and State University, Blacksburg, Virginia 24061, USA}
\affiliation{Department of Chemistry, Virginia Polytechnic Institute and State University, Blacksburg, Virginia 24061, USA}
\author{John J. Lesko}
\affiliation{Macromolecules Innovation Institute, Virginia Polytechnic Institute and State University, Blacksburg, Virginia 24061, USA}
\affiliation{Department of Mechanical Engineering, Virginia Polytechnic Institute and State University, Blacksburg, Virginia 24061, USA}
%\email{jlesko@vt.edu}

\date{\today}

\begin{abstract}
\par We use molecular dynamics simulations to probe hydration, ion spacing, and cation-anion interaction in two sulfonated polysulfones with different ion distributions along the polymer backbone. At room temperature, these polymers are below their experimental glass transition temperatures even with water contents more than $10\%$. At the equilibrium water uptake, the ions exhibit a similar level of hydration as they would in their saturated aqueous solution. The framework of Manning's limiting law for counterion condensation is used to examine ionic interactions in the simulated polysulfones. The dielectric constant ($\varepsilon$) that the ions experience can be well approximated by a volume-weighted average of the dielectric constants of the polymer backbone and water. Our results show that a reasonable estimate of the average inter-ion distance, $b$, is obtained by using the distance where the sulfonate-sulfonate coordination number reaches 1. The spacing of the sulfonate ions along the polysulfone backbone plays a role in determining their spatial distribution inside the hydrated polymer. As a result, the value of $b$ is slightly larger for polymers where the sulfonate ions are more evenly spaced along the backbone, which is consistent with experimental evidence. The simulations reveal that the sulfonate ions and sodium counterions form fibrillar aggregates at water contents below the equilibrium water uptake. Such extensive ionic aggregates are expected to facilitate ion transport in sulfonated polysulfone membranes, without the need for long-range chain motion as in the case of traditional rubbery ionic polymers. Our estimates for $\varepsilon$ and $b$ are used in conjunction with Manning's theory to estimate the fraction of counterions condensed to the fixed ions. The prediction of Manning's theory agrees well with the result computed by directly counting the condensed sodium ions in the molecular dynamics simulations.
\end{abstract}

\maketitle

\section{Introduction}

\par Polymeric membranes based on ionic polysulfones have been suggested for a number of applications including fuel cells,\cite{Harrison2005} ionic actuators,\cite{Tang2014} water electrolysis,\cite{daryaei2017-1} reverse osmosis,\cite{daryaei2017-2} and electrodialysis.\cite{choudhury2019} All of these applications rely on the interplay between the hydration of the polymer, the mobility of the counterions, and the exclusion of the coions in order for the ionic polymer to perform a specific function. However, our fundamental understanding of how the chemical structure of a glassy ionic polymer dictates its bulk properties, such as water uptake, ion transport, coion exclusion, electrical resistance, and thermomechanical response is limited. More specifically, the bulk properties of a glassy ionic polymer often scale with its ion content only over a limited range, and the physical origins of the apparent transitions from one scaling regime to another are poorly understood.\cite{daryaei2017-1, daryaei2017-2, choudhury2019} In this paper, we use molecular dynamics (MD) simulations to investigate the interrelated effects of hydration, fixed ion spacing, and cation-anion interaction in sulfonated polysulfones in order to gain insight into the molecular-scale processes, such as ionic aggregation and counterion condensation, that underpin their bulk properties. Insight at the molecular level will provide a foundation for future work exploring the bulk properties of ionic polysulfone membranes including equilibrium hydration, density, mechanical performance, ion exclusion, and ion transport.

\par The prevailing framework for understanding ionic interactions in polyelectrolyte solutions is the counterion condensation theory developed by Manning.\cite{manning_limiting_1969} This theory was originally conceived to describe the behavior of rod-like polyelectrolytes in a dilute solution. It can be regarded as an extension of the Debye-H\"uckel limiting law. The parameter that Manning proposed to quantify the effect of counterion condensation is
\begin{equation}
    \xi=\frac{l_\text{B}}{b}=\frac{e^2}{4\pi\varepsilon_0\varepsilon k_\text{B} T b}~,
    \label{eq:Manning}
\end{equation}
where $l_\text{B}$ is the Bjerrum length, $b$ is the average distance between the fixed ions, $e$ is the elementary charge, $\varepsilon_0$ is the vacuum permittivity, $\varepsilon$ is the dielectric constant of the medium that solvates the ions, $k_\text{B}$ is the Boltzmann constant, and $T$ is the absolute temperature of the system. The valences of the fixed and mobile ions are denoted $z_i$ and $z_p$, respectively. In Manning's model, if $\xi>1/(z_i z_p)$, then counterion condensation occurs. For the sulfonated polysulfones studied here, both fixed ions and counterions are monovalent, so $z_i=z_p=1$. Therefore, the critical value of the Manning parameter, $\xi_\text{c}$, is 1. When $\xi> \xi_\text{c}$, the mobile counterions will condense onto the fixed ions. Since the condensed ions are ``neutralized" from an electrostatic point of view, the effective Manning parameter will adopt the critical value ($\xi_\text{c}$) when calculated using only the uncondensed counterions. The fraction of the condensed ions to all the ions in the system can then be expressed as
\begin{equation}
    f_\text{c}=1-\xi_\text{c}/\xi~,
    \label{eq:condensedfrac}
\end{equation}
which indicates that there are no condensed ions when $\xi = \xi_\text{c}$ and the number of condensed ions increases as $\xi$ increases beyond $\xi_\text{c}$. Since only the uncondensed ions contribute to the Donnan potential of the system, the ability of a membrane to exclude coions is directly related $f_\text{c}$.\cite{manning_limiting_1969}

\par Because of its relevance to membrane selectivity, Manning's counterion condensation theory has been extended to dense ionic polymer membranes.\cite{kamcev_effect_2017} We take the same approach here. We set $T=300$ K because for most applications membranes are at or near room temperature. Since $\varepsilon_0$ and $k_\text{B}$ are both constants, $\varepsilon$ and $b$ remain as characteristic variables of a specific system. In the dilute solution limit that Manning originally investigated, the determination of $\varepsilon$ and $b$ is straightforward. The contribution of the polymer backbone in a dilute solution to its dielectric constant is negligible and $\varepsilon$ is just the dielectric constant of the solvent, usually water. In the dilute limit, charged polymers adopt rod-like conformations and the average linear distance between fixed ions along the backbone provides a good estimate for $b$. 

\par It is much less clear how $\varepsilon$ and $b$ should be determined for concentrated solutions or solid polymers, where the ions exist within a mixture of solvent molecules and polymer backbones. Conceptually, $\varepsilon$ must represent a combined polymer-solvent dielectric constant consistent with the ion's environment.\cite{kamcev_effect_2017} The average spacing of the fixed ions, $b$, presents even more of a challenge in the concentrated regime. If the ions are evenly distributed within the hydrated polymer, then $b$ should be inversely related to the average ion density. However, dense ionic polymers are often phase separated, which means that the ion-rich phase can have a local ion density much higher than the average value, resulting in a value for $b$ smaller than that computed with the average ion density. Furthermore, in some polymers the ions are paired or grouped along the backbones, in which case the local value of $b$ within an ion cluster is different from the average value of $b$ along a backbone. Since it is difficult to measure $b$ experimentally, especially for sodium-sulfonate systems where the contrast is low for X-ray scattering, we turn to MD simulations in order to understand the molecular-scale behavior of ionic polysulfones in a dense glassy state.

\par Within the past few decades, advances in computational power have enabled:
\begin{enumerate}
    \item More robust quantum mechanical methods, which have been used to improve our understanding of the hydrogen bond network of water around small solute molecules.\cite{varma_2006,rowley2012, wang2013, shiau_cluster-continuum_2013, shishlov_2016}
    \item Larger-scale MD simulations, which have made it possible to simulate hydrated macromolecules.\cite{Charkhesht2018, tamai1996-1, tamai1996}
    \item Improved signal processing and data acquisition rates, which have made experimental characterization methods for liquid water's molecular dynamics, such as neutron diffraction and microwave and terahertz spectroscopy, more accessible.\cite{blake_microwave_2001, mancinelli_2007, Charkhesht2019, ben-amotz_2019}
\end{enumerate}
Since both simulation and experimental methods have made significant progress, the current challenge lies in finding ways for simulations to deepen our understanding of the physicochemical mechanisms underlying experimental observations. The goal here is to use atomistic simulations to better understand the trends in experimental data and the relationships between water content, ion content, and backbone chemistry. A deeper understanding of these relationships will enable us to apply systematic design principles in the development of solid ionic polymers for membrane applications. Such an approach could allow us to more fully optimize the polymer architecture for particular applications, thereby improving the performance of ionic polymer membranes.

\par In the past, all-atom MD simulations have been applied to study polysulfones. One early work by Fan and Hsu indicated that the barrier to rotation is the lowest for the C-S bond, rather than the C-O bond, in a polysuflone backbone.\cite{fan_application_1991} Marque \textit{et al.} investigated the interactions between the polar groups in a non-ionic polysulfone backbone and water.\cite{marque2010} Ionic polysulfones have also been studied with atomistic simulations in recent years. Merinov and Goddard simulated a polysulfone with quaternary ammonium cations on its backbone and OH$^-$ anions in both dry and hydrated state.\cite{merinov2013} Han \textit{et al.} simulated phase-separated polysulfones with either quaternary ammonium-functionalized fluorinated side groups or sulfonated fluorinated side groups.\cite{han2014} They found that the concentration contrast between the phases is less distinct in the sulfonated polymer.\cite{han2014} Wohlfarth \textit{et al.} conducted a thorough investigation of the proton-association phenomena in sulfonated and hypersulfonated polysulfones.\cite{wohlfarth_proton_2015} They found evidence for proton sharing between the sulfonate groups and discussed the practical implications of the counterion condensation phenomena for ion conduction.\cite{wohlfarth_proton_2015}

\par Other researchers have also used all-atom MD simulations to investigate ion distribution and aggregation in sulfonated polymers. Notably, Lin and Maranas found evidence of chain-like ionic aggregates in a sulfonated polyethylene oxide-based ionomer.\cite{lin_cation_2012} More recently, Abbott and Frischknecht used atomistic simulations to investigate percolating ionic domains in sulfonated polyphenylenes.\cite{abbott_nanoscale_2017} There has also been increased interest in developing theories for ionic aggregation.\cite{choi_graph_2018, mceldrew_theory_2020} Moreover, some researchers including Bahlakeh \textit{et al.} analyzed the structure of water aggregates within a hydrated polymer.\cite{bahlakeh2013}

\par In this paper, we use MD simulations to gain a deeper understanding of the spatial distribution of ions within sulfonated polysulfones at various levels of hydration, with sodium ions as the counterions. Other cations such as $\text{Li}^+$, $\text{K}^+$, $\text{Rb}^+$, and $\text{Cs}^+$ exhibit different hydration characteristics. A change in cation will thus change the equilibrium water uptake of the polymer, which is of future interest. The polysulfones in the sodium-ion form studied here remain in a glassy state even at room temperature and ambient moisture contents greater than $10\%$. Based on Manning's counterion condensation theory, the inter-ion distance is a central parameter in determining the utility of ionic polymers for specific membrane applications. While we have a more complete picture of phase separation and inter-ion spacing in hydrated Nafion and other rubbery polymers,\cite{jang2004, chou2005, venkatnathan2007, lu2008} our understanding of the ion distribution in hydrated polysulfones and other glassy ionic polymers with aromatic backbones is less robust. Although ionic aggregates were observed in some glassy ionic polymers, \cite{chang_effect_2015,weiber_highly_2013,sorte_impact_2019} there is no experimental evidence for the formation of large scale (microscale) ionic domains in the polysulfones investigated here, which leads to the assumption that the ions are evenly distributed within these hydrated polymers.\cite{sundell2014} However, relying on this assumption limits our ability to apply systematic design principles to the development of new aromatic polymers. A more thorough understanding of ion spacing and distribution in polysulfones will therefore enable us to rationally design new polymers with tailored physical properties desired in specific applications.\cite{kocherbitov2010}

\section{Materials and Methods}

\subsection{Materials}

\begin{figure*}[htb]
   \centering
   \includegraphics[width=\textwidth]{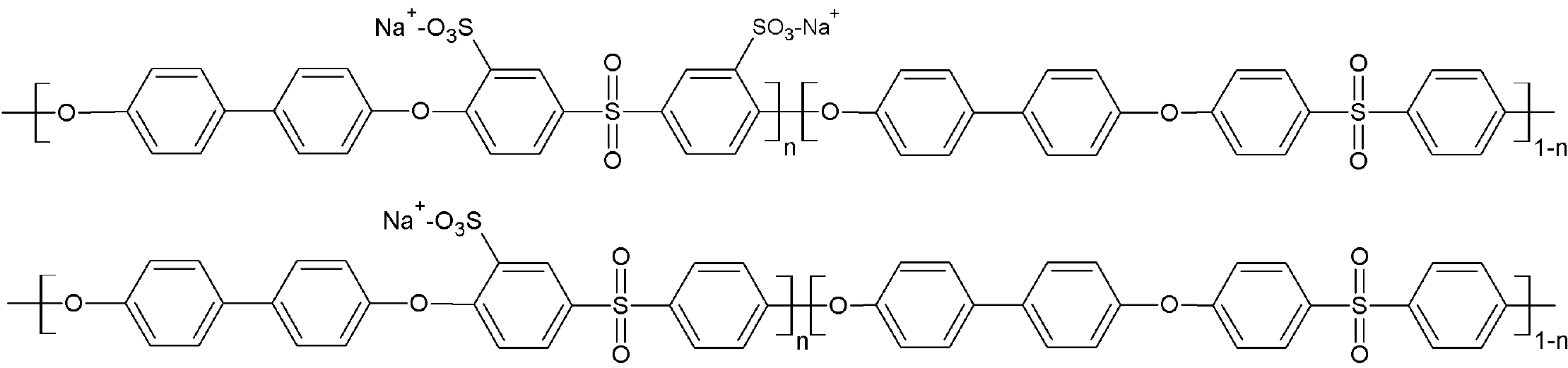}
   \caption{Chemical structures of the sulfonated polysulfones: dBPS - disulfonated (top) and mBPS - monosulfonated (bottom).}
   \label{fig:structure_exp}
\end{figure*}

\par The chemical structures of the sulfonated polysulfones that we investigate here are shown in Fig.~\ref{fig:structure_exp} and an overview of their properties is presented in Table~\ref{TB1}. The polymers are synthesized from pre-sulfonated sulfone monomers and exchanged to their sodium counterion form as described elsewhere.\cite{daryaei2017Diss, Wang2001} The only difference between the two polymers is the spacing of the sulfonate ions along the polymer backbone. In the disulfonated polymer, two sulfonate ions are always located on the same monomer while in the monosulfonated polymer, there is only one ion per monomer. We refer to a monosulfonated (disulfonated) polysulfone as mBPS$x$ (dBPS$x$), where $x$\% is the fraction of the sulfone monomers that are sulfonated expressed as a percent. The content of fixed ions is defined as the ratio of the number of sulfonate groups to the number of sulfone groups on all the polymer chains. Therefore, the ion content of polymer mBPS$x$ is $x$\% while for dBPS$x$, it is $2x$\%. The polymers of greatest interest for desalination applications have ion contents near 50\%, so we use tetramers sulfonated at 50\% for our simulations. Since the dBPS25 polymer was not available for experiment, we present the experimental data for the dBPS22 and dBPS27 polymers instead. Details of the experimental procedures used to obtain the data in Table~\ref{TB1} are included in the Supporting Information.

\begin{table*}[htb]
	\centering
	\caption{Experimentally measured properties of the sulfonated polysulfones in their sodium salt form.}
		\begin{tabular}{|c|c|c|c|c|c|c|c|}
	    \hline
		Polymer & Molecular & Ion & IEC & $f_\text{wu}$  & $\lambda_\text{eq}$ & Hydrated & Density\\
		        & Weight (kDa)    & Content &     & (\%)   &    & $T_g$ (\degree C) & (g/cm$^3$)\\
		\hline     
BPS0 & 120         &  0 & 0    &2.6     & 0 & --     & 1.33  \\ \hline
mBPS50 & 127       & 50 & 1.16 &16~$\pm$~1  &  7.7~$\pm$~0.5 & 157  & 1.34  \\ \hline
dBPS22 & 97        & 44 &0.99 & 14~$\pm$~3  &  7.6~$\pm$~2  & 177  & 1.34 \\ \hline
dBPS27 & 144       & 54 &1.18 & 18~$\pm$~2  & 8.6~$\pm$~0.9  & 169 & 1.35  \\ \hline
\end{tabular}
\label{TB1}
\end{table*}

\subsection{Quantum Chemical Calculations}
\label{Quantum}

\par
\begin{figure*}[htb]
   \centering
   \includegraphics[width=\textwidth]{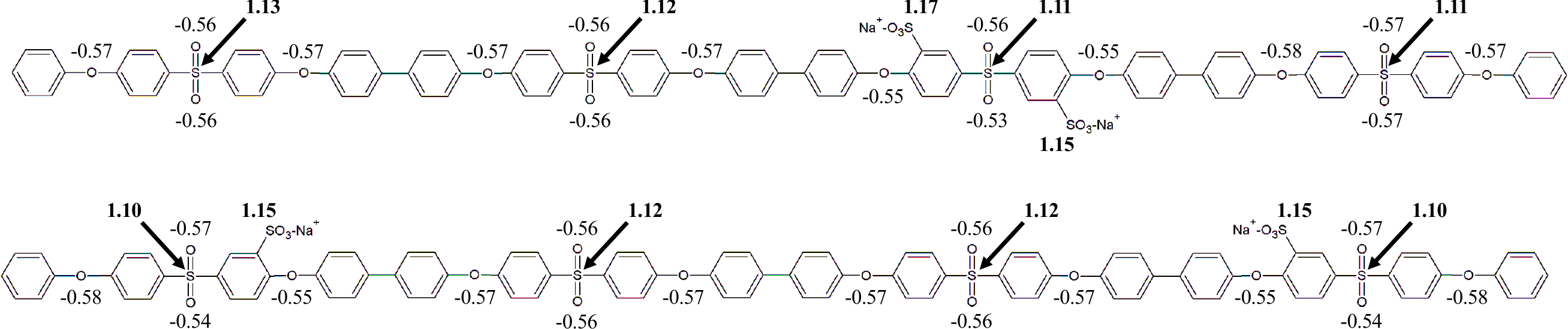}
   \caption{Structures of the model oligomers: disulfonated (top, representing dBPS25) and monosulfonated (bottom, representing mBPS50) tetramer. The partial charges of oxygen atoms in the sulfone groups and ether linkages as well as those of all sulfur atoms are included in the figure. Other partial charges (average values): carbon atoms adjacent to an ether linkage: 0.34; all other carbon atoms: -0.14; oxygen atoms in the sulfonate groups in the monosulfonated oligomers: -0.59; oxygen atoms in the sulfonate groups in the disulfonated oligomers: -0.60; hydrogen atoms: 0.15. All charge values are reported in the unit of the elementary charge.} 
   \label{fig:structure_MD}
\end{figure*}

 Representative tetramers of the sulfonated polysulfones are built using the MAPS molecular builder.\cite{maps} Each model oligomer contains four sulfone units, three biphenol units, and two phenol end groups as shown in Fig.~\ref{fig:structure_MD}. We choose to simulate an ion content of two sulfonates per sulfone tetramer because it represents the ion content of most interest for desalination applications. For the disulfonated tetramer, one of the central sulfone units is disulfonated in order to minimize interference from the end groups. We call this tetramer dBPS25 because 25\% of the monomers are disulfonated. For the monosulfonated tetramer, the two sulfonate ions are placed on the two end sulfone units in order to minimize their mutual interaction. We call this tetramer mBPS50 because 50\% of the monomers are monosulfonated. Both mBPS25 and dPBS50 tetramers used in the simulations contain 2 sulfone monomers per fixed ion (i.e., 0.5 fixed ion per monomer) and therefore are directly comparable. The use of tetramers allows us to model a sufficiently large system of which the interactions and aggregation behavior of ions can be captured. However, the effects of chain conformations, such as loop formation and chain ordering, on ionic aggregation may not be fully treated in the tetramer model because of the relatively short chains. Such effects can be explored with future simulations of longer chains.

\par We use GAUSSIAN09 software to perform quantum chemical calculations to obtain the partial charges of all atoms in the model oligomers.\cite{g16} The atomic structures of the model oligomers built with the MAPS molecular builder are used as initial input. Sodium counterions are not included in the optimization and the total charge of the oligomer is thus set to -2 in the unit of the elementary charge. The system is first optimized using the semi-empirical PM6 method. The output from the PM6 method is used as input for the final optimization using the B3LYP method with the 6-31G(p) basis set. Statistically similar values are obtained using the 6-31G(d,p) basis set. 

\par The partial charges resulting from the quantum chemical calculations are included in Fig.~\ref{fig:structure_MD} and its caption. These charges are assigned to the atoms of the tetrameric polysulfones and used to compute the  electrostatic interactions in the MD simulations described below. The charges for the backbone atoms from our calculations are very similar to those reported in the literature for non-sulfonated polysulfones\cite{marque2010} and the charges for the atoms in the sulfonate groups are also similar to those reported for other sulfonated polymers.\cite{hu2006}

\subsection{Molecular Dynamics Simulations}
\label{MD}

\par We perform classical MD simulations on systems consisting of 64 sulfonated sulfone tetramers. The tetramers are built as described in the previous section. The resulting 256 sulfone monomers are 50\% sulfonated, which means that each system contains 128 fixed sulfonate ions and 128 sodium counterions. Since $\lambda$ indicates the number of water molecules per cation-anion pair, the total number of water molecules in the system is $128\lambda$. Table \ref{TB1} shows the experimental values of $\lambda_\text{eq}$ for the polymers of interest. For the 50\% sulfonated polymers, $\lambda_\text{eq} \simeq 8$, meaning that there are around 8 water molecules per ion pair. In our simulations, $\lambda$ is varied from 3 to 14. Water contents greater than $\lambda_\text{eq}$ are not experimentally attainable at room temperature. However, with simulations we can investigate water contents higher than $\lambda_\text{eq}$. This helps us establish trends and understand the significance of high water contents. It may also shed light on the behavior of polymers with higher ion contents, even though our simulations are conducted at a single fixed ion content.

\par MD simulations are carried out using LAMMPS\cite{Plimpton1995} with the PCFF force field.\cite{sun1994} A PCFF water model is adopted with a charge 0.3991$e$ on hydrogen and -0.7982$e$ on oxygen, which are computed using MAPS with the bond-increment method. These values are almost identical with those reported by Consiglio and Forte for PCFF water.\cite{consiglio_pcff_water_2018} The charges of the atoms on the polysulfone chains are determined with the quantum chemical calculations discussed in the previous section. These charges are used for computing the electrostatic interactions. The equations of motions are integrated with a velocity-Verlet algorithm with a time step of 1 fs. The temperature is controlled using a Nos\'e-Hoover thermostat. A dry polymer system is first built using 64 tetramers and equilibrated by thermal cycling. Then an appropriate number of water molecules are randomly inserted into the simulation box in order to obtain the target value of $\lambda$. The total number of atoms in the simulations ranges from 13312 for $\lambda=3$ to 17536 for $\lambda=14$. The resulting rectangular simulation boxes have side lengths between 5.4 and 5.8 nm, with periodic boundary conditions imposed along each Cartesian axis. After adding water molecules, the simulations are run in an isothermal–isobaric ensemble (NPT) at 1 atm and 300K for 5 ns in order to accommodate the water molecules. After 5 ns, the density of the hydrated polymer has stabilized, and then the system is switched to a canonical ensemble (NVT) and heated to 600K over 2 ns. The glass transition temperature of the polymer is well below 600K (see Table \ref{TB1}), so this heating step provides sufficient backbone motion for the system to reach a low energy state consistent with the state of the polymer in the solution cast films used for experiment. After that, the system is maintained at 600K for 2 ns to further equilibrate and cooled down to 300K over the subsequent 2 ns period. Finally, a data collection run is conducted in NPT at 300K and 1 atm for 2 ns. 

\par Snapshots are collected every 1,000 time steps, resulting in 2,000 snapshots for each system, of which the last 1,000 (i.e., the data collected in the last 1 ns) are fully analyzed to generate the data reported herein. In general, the error bars on the plots represent the standard deviation of the plotted values over the 1,000 snapshots that are analyzed. To further improve the accuracy of the data on the inter-sulfonate distances, we heat up the hydrated polymers at $\lambda = 4$, 8, and 12 and construct three new starting configurations for these polymers. Then the cooling, reequilibration, and data collection procedures described above are repeated for each new starting configuration. As a result, there are four independent simulations with different starting configurations. The results in Fig.~\ref{fig:SO3SO3RDF} below are obtained by averaging over these four simulations.

\par Based on MD simulation output, which consists of atom coordinates, we primarily use a radial distribution function (RDF) to analyze ion hydration, spatial distribution of ionic groups, and cation-anion interaction. The RDF represents the local density of a particular type of atoms of interest around a central atom and is denoted $g(r)$, which can be computed as
\begin{equation}
    g(r)=\frac{N}{\rho_0 4 \pi r^2 \text{d}r}~,
    \label{eq:gr}
\end{equation}
where $r$ is the distance from the central atom, $N$ is the number of the atoms of interest within a spherical shell from $r$ to $r+\text{d}r$ around the central atom, and $\rho_0$ is the average number density of the atoms of interest in the system. In computing $g(r)$ for hydrated systems, it is typical to regard the position of the oxygen atom as the location of a water molecule.\cite{Biman_2013} For a sulfonate group, the position of the central sulfur atom will be used as its location. The number of atoms within the first peak in the RDF around an ion is termed the coordination number, denoted $C_n$. Mathematically,
\begin{equation}
    C_n = \rho_0 \int_0^{r_\text{min}} g(r) 4\pi r^2 \text{d}r~,
    \label{eq:coord_number}
\end{equation}
where $r_\text{min}$ is the location of the first trough in $g(r)$.

\begin{figure}[htb]
   \centering
   \includegraphics[width=0.45\textwidth]{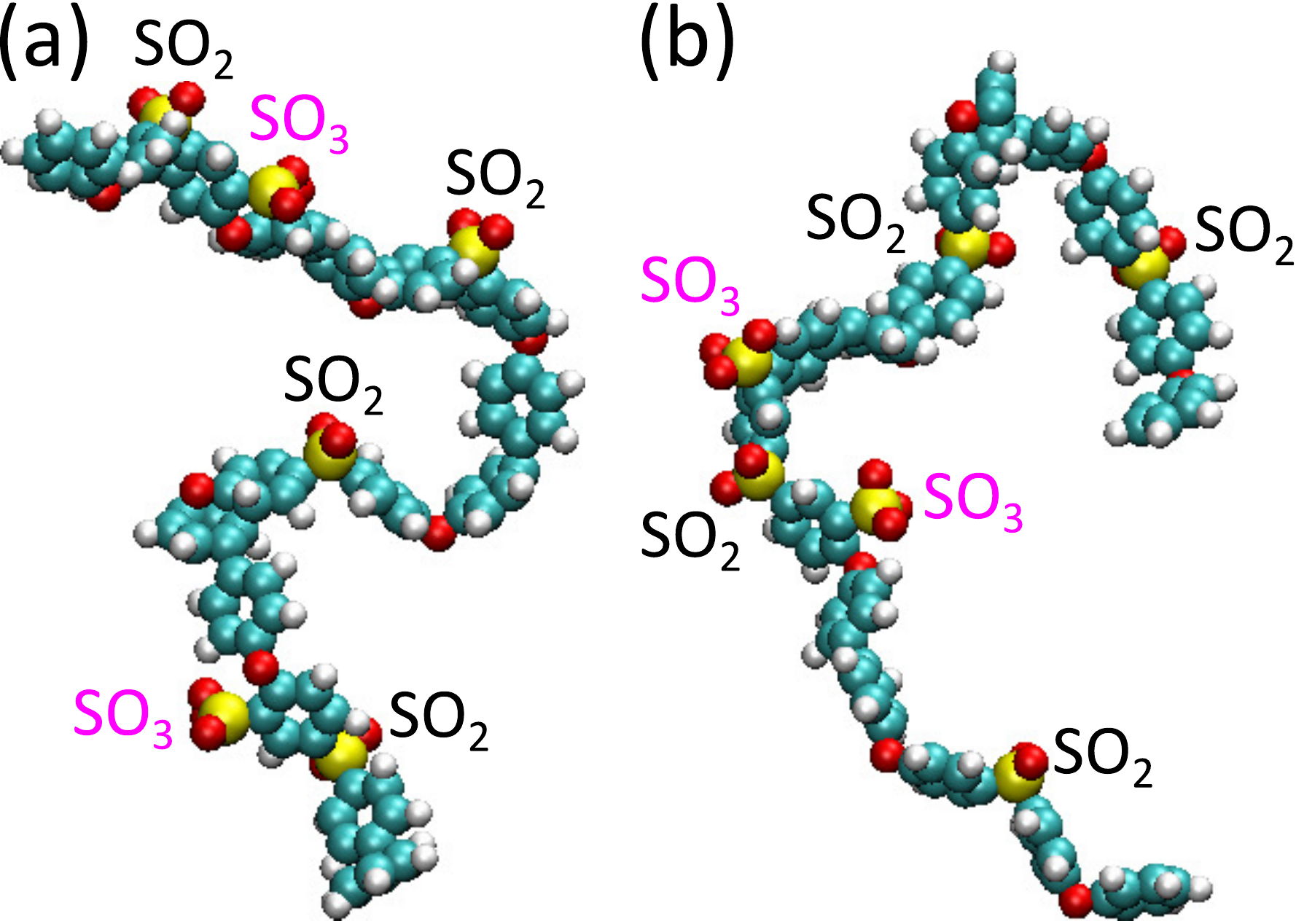}
   \caption{Representative conformations of the (a) mBPS50 and (b) dBPS25 tetramers from MD simulations.}
   \label{fig:gauss_view}
\end{figure}

\section{Results and Discussion}

\par
\begin{figure*}[htb]
   \centering
   \includegraphics[width=\textwidth]{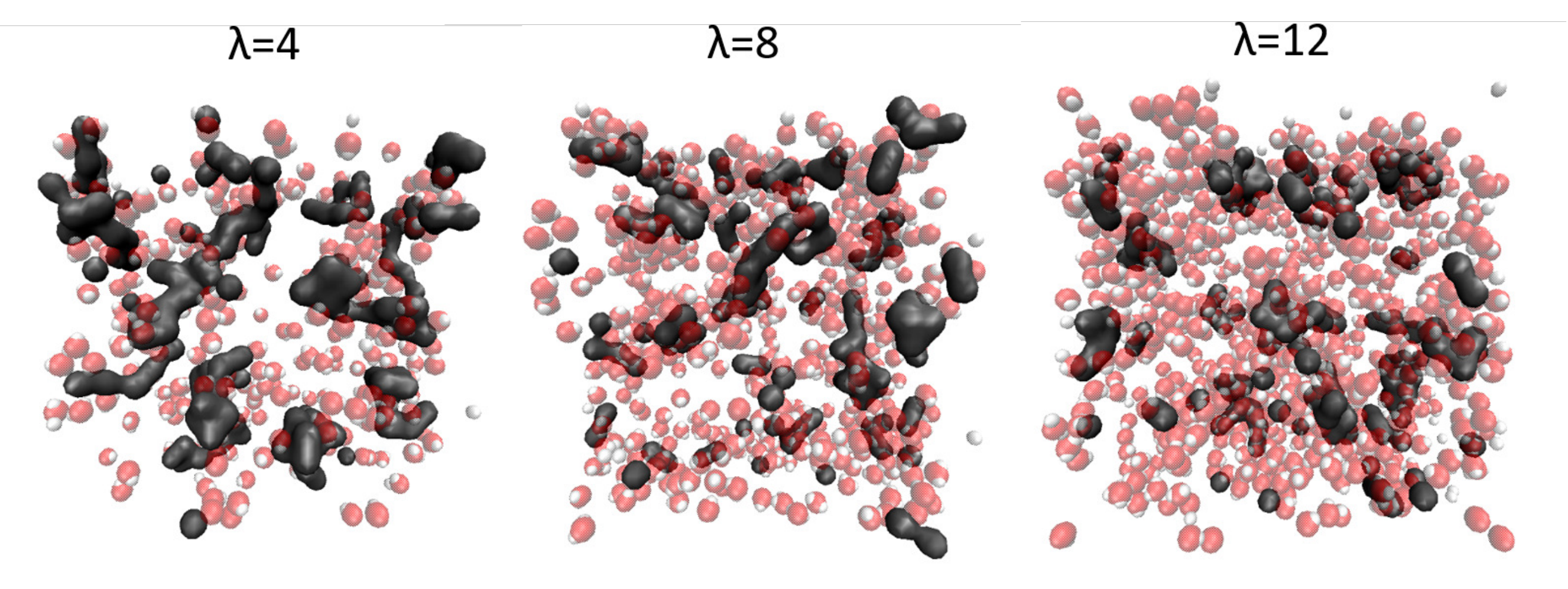}
   \caption{Snapshots of the disulfonated tetramers at $\lambda=4$, $8$, and $12$. In each snapshot, a slice of 20~\AA~ thick is shown. Ionic aggregates are shown as dark grey surfaces. Oxygen and hydrogen atoms of water molecules are shown in transparent red and white, respectively. Backbone atoms are not visualized to improve clarity.} 
   \label{fig:MD_snapshots}
\end{figure*}

 Two representative tetramer conformations from MD simulations are shown in Fig.~\ref{fig:gauss_view}. Three snapshots of the hydrated disulfonated tetramers at $\lambda=4$, 8, and 12 are shown in Fig.~\ref{fig:MD_snapshots}. For the $\lambda=8$ case, which is most comparable to the hydration level in the experiments, the equilibrium density from MD simulations is 1.21 g/cm$^3$ for the monosulfonated polymer and 1.22 g/cm$^3$ for the disulfonated one, respectively. These densities are less than 10\% lower than the experimental hydrated densities shown in Table \ref{TB1} for much longer chains. The simulation data show that the disulfonated polymer is slightly denser than the monosulfonated polymer, which also agrees well with experimental measurements. Based on the snapshots from simulations, the ions appear to be dispersed throughout a hydrated polysulfone and there are not noticeable ion-rich or polymer-rich regions. The apparent lack of large-scale ionic aggregates is in good agreement with the solid state $^{23}$Na NMR and SAXS results for similar polymers.\cite{sundell2014} The aggregation behavior of ions will be discussed in much more detail in Section \ref{iondist}.

\par In the following sections, we report the simulation results on hydration, ion distribution, and cation-anion interaction (i.e., ion aggregation and counterion condensation). We compare the results of the monosulfonated and disulfonated polysulfones to gain insight into the differences that arise due to the difference in ion distribution along the backbone. In Sec.~\ref{hydration}, we examine the extent to which the ions are hydrated at different water contents. This analysis provides insight into the estimation of the combined dielectric constant ($\varepsilon$) that should be applied to the sulfonated polysulfones. Next, in Sec.~\ref{iondist}, we examine the spatial distribution of the sulfonate ions within the hydrated polymers and investigate how the inter-sulfonate distance differs between the two polymers of interest. We discuss methods to determine the average inter-ion distance, $b$. Finally, in Sec.~\ref{pairing} we use the estimates for $\varepsilon$ and $b$ to more fully understand the anion-cation interaction and discuss their implications for the ion transport through the ionic polysulfone membranes. Overall, the simulation results provide an atomic-scale picture of how the distribution of ions along an aromatic polymer backbone is related to hydration, ion aggregation, coion exclusion, and counterion transport in polysulfone-based membranes.

\subsection{Hydration of Ions}
\label{hydration}

\par In general, our simulations (e.g., the snapshots in Fig.~\ref{fig:MD_snapshots}) show that in the sulfonated polysulfones, the ionic regions are hydrated, but many water molecules are not within the primary hydration shells of the ions. In the following, we investigate ion hydration more quantitatively. We first examine the coordination number of water molecules around ions to understand whether the ions are strongly or weakly hydrated compared to those in aqueous solutions. Then, we quantify the fraction of water molecules that are directly associated with the ions at different water contents and discuss the implications of this result for the dielectric constant ($\varepsilon$) that the ions experience within the hydrated polymer.

\begin{figure}[htb]
    \centering
    \includegraphics[width=.45\textwidth]{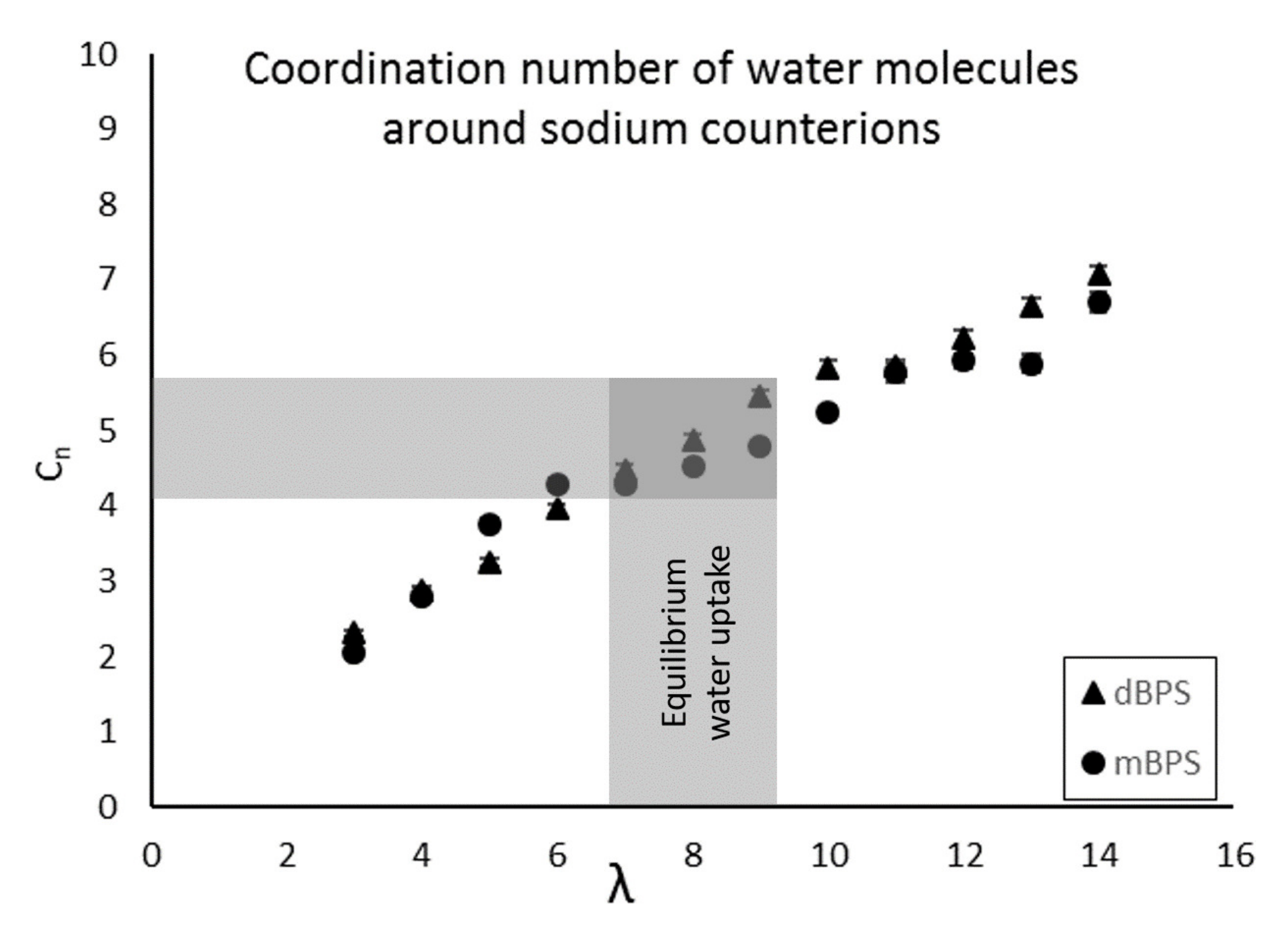}
    \caption{Coordination number of water around the sodium counterion for dBPS25 (triangles) and mBPS50 (circles). Cutoff radius is 4.55~\AA.} 
    \label{fig:NaWcoord}
    \includegraphics[width=.45\textwidth]{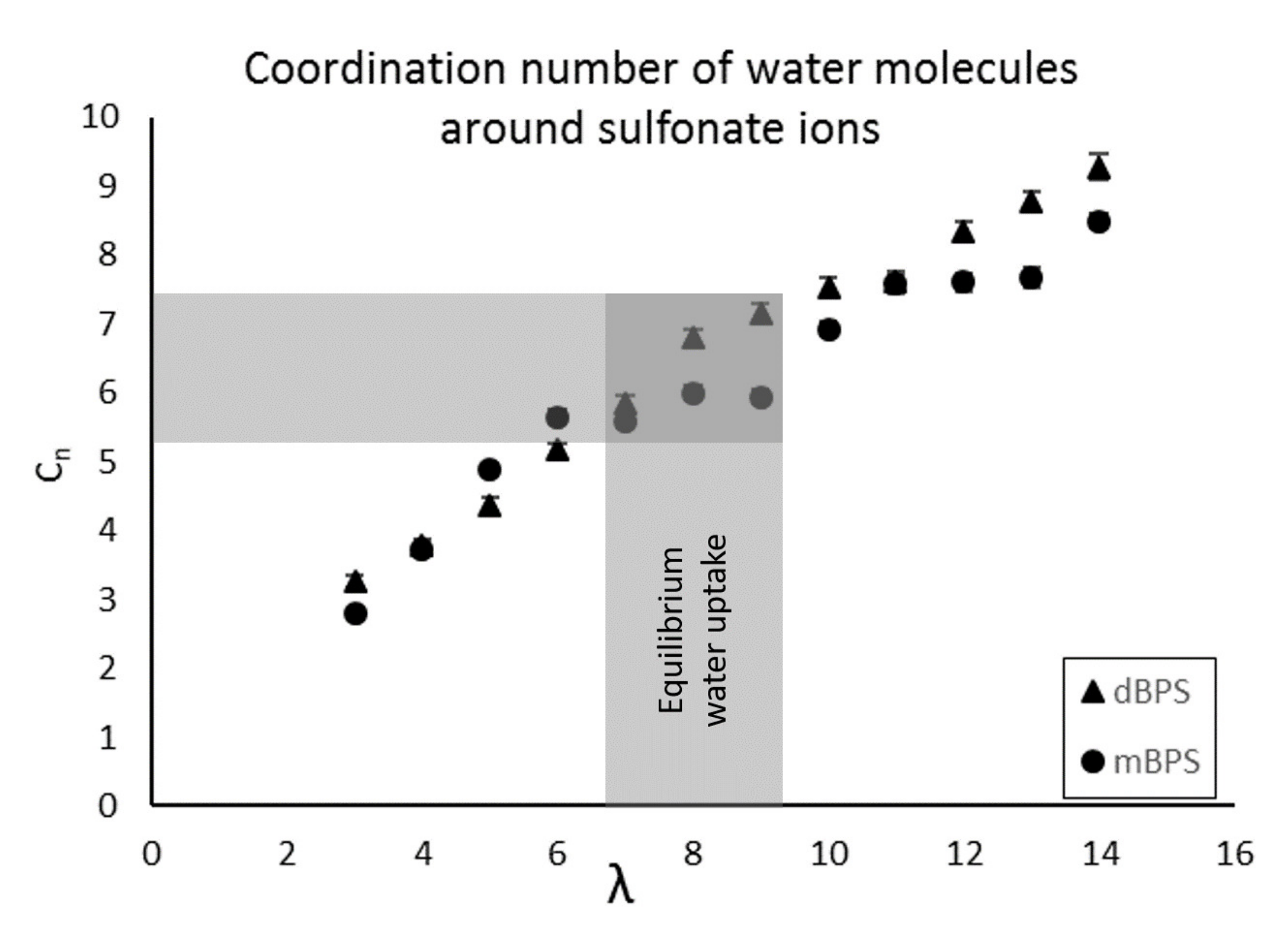}
    \caption{Coordination number of water around the sulfonate group for dBPS25 (triangles) and mBPS50 (circles). Cutoff radius is 5.45 \AA} 
    \label{fig:SO3Wcoord}
\end{figure}

The coordination number of water, $C_n$, around $\text{Na}^+$ and $\text{SO}_3^-$ ions is shown in Figs.~\ref{fig:NaWcoord} and \ref{fig:SO3Wcoord}. As expected, $C_n$ increases as the water content is increased. For $\lambda < \lambda_\text{eq}$, both monosulfonated and disulfonated polymers exhibit similar hydration behavior. At the equilibrium water uptake, $C_n \simeq 4.5$ for $\text{Na}^+$, which is consistent with the coordination number of $\text{Na}^+$ at the limit of its solubility in water.\cite{mancinelli_2007} For $\text{SO}_3^-$, $C_n \simeq 6$ at the equilibrium water uptake, which indicates 2 hydrogen bonds per sulfonate oxygen. This is consistent with the minimum water coordination number for the solubility of sulfonates.\cite{venkatnathan2007, vchirawongkwin2012} We conclude that the equilibrium water uptake of the polymer is sufficient to provide the minimum solvation shell for the ions, but no more. In this way, the polymer at its equilibrium water uptake shares some physical characteristics with a saturated aqueous solution of the ions involved. Therefore, it should be reasonable to apply the theory of ionic aggregation in saturated solutions to solid-state ionic polymers.

\par Although experimentally unattainable, we can still probe the high water uptake regime with MD simulations. The data in Figs.~\ref{fig:NaWcoord} and \ref{fig:SO3Wcoord} show that when $\lambda > \lambda_\text{eq}$, a statistically significant difference in $C_n$ appears to emerge between the monosulfonated and disulfonated polymers. For both $\text{Na}^+$ and $\text{SO}_3^-$ ions, $C_n$ is higher in the disulfonated case, indicating a slightly higher level of hydration of the ions in the disulfonated polymer. Experimental evidence also suggests that at high ion contents, the equilibrium water uptake of the disulfonated polymer is larger than that of the monosulfonated polymer at the same ion content.\cite{daryaei2017Diss} Therefore, the simulation and experiment are consistent in this respect. However, the mechanism underlying this difference in hydration remains unclear.

\begin{figure}[htb]
    \centering
    \includegraphics[width=.45\textwidth]{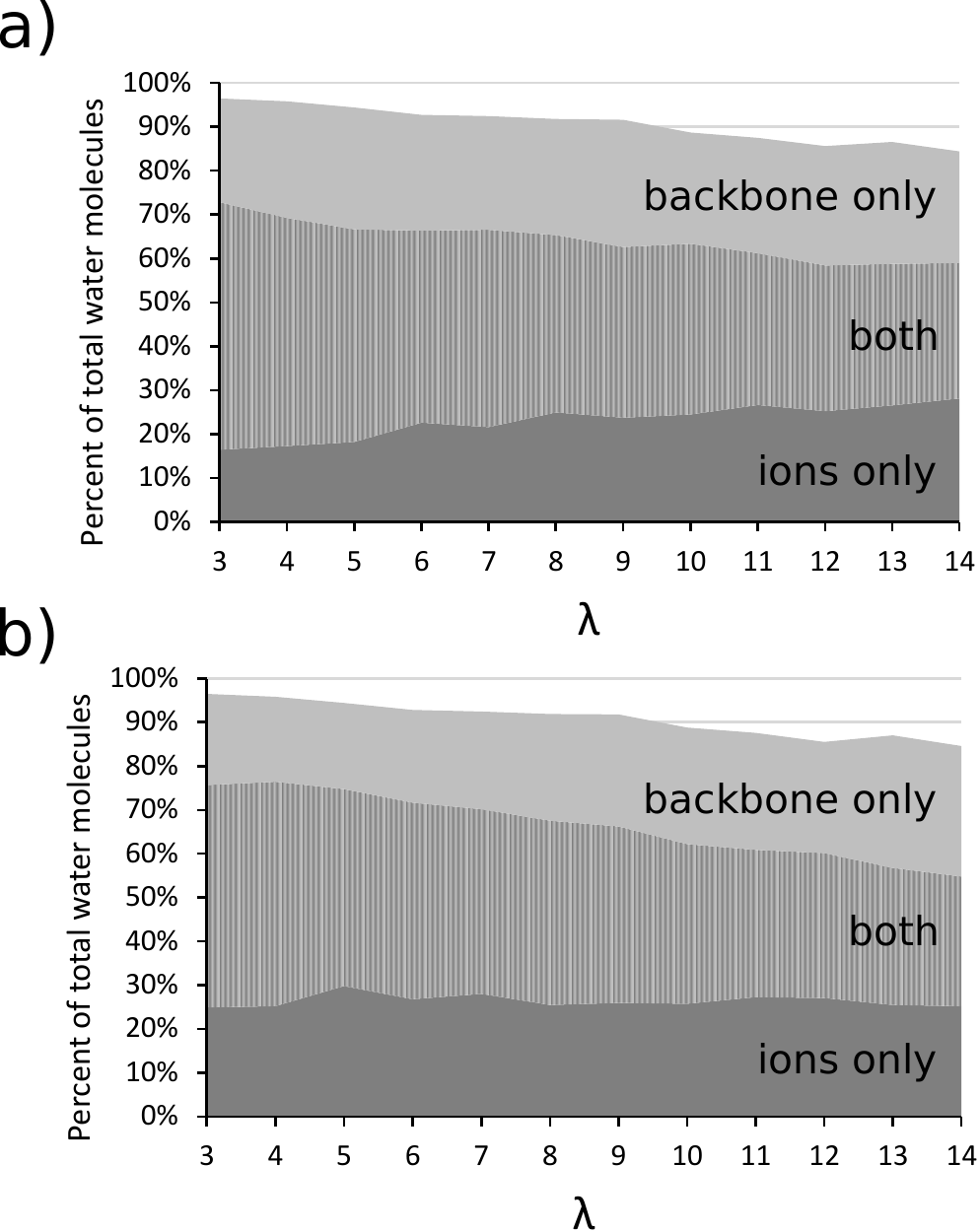}
    \caption{Stacked area plots showing the fraction of water molecules that is associated with the ions, the backbone polar groups (sulfone and ether linkages), and both, for the a) mBPS50 and b) dBPS25 tetramers. Cutoff distances to determine the associated water are set at the first minimum in the corresponding RDF. The white space at the top of each plot represents the fraction of water not associated with either an ion or a polar group.} 
    \label{fig:waterstacks}
\end{figure}

\par Since a water molecule can form 4 hydrogen bonds, they can be shared between multiple ions. This phenomenon can be seen directly in the snapshots shown in Fig.~\ref{fig:MD_snapshots}, where the primary hydration shells of two or three adjacent ions are found to overlap. There are also quite a few water molecules not within the primary hydration shells of any ions at all. Here we compute the fraction of water molecules that is associated to the ions and polar groups. The results are shown as the stacked area plots in Fig.\ref{fig:waterstacks}. Based on these plots, 55\% to 75\% of the water molecules are ion-associated. This means that at low water contents, even though the ions may be weakly hydrated, at least a quarter of the water molecules are located away from the ions. These water molecules are mostly near the polar groups of the backbone. Upon closer inspection of Fig.\ref{fig:structure_MD}, the partial charge on the sulfonate oxygen atoms is only slightly larger than that on the oxygen atoms of the ether and sulfone linkages. This implies that the backbone plays a significant role in the hydration behavior of polysulfones. This result is consistent with the experimental observation that the polarity of the backbone has a significant effect on water uptake and other bulk properties of polysulfones.\cite{choudhury2019}

\par Taken all together, the simulations indicate that water does not have a particularly strong preference for the ions. In this sense, the sulfonated polysulfones are in contrast to phase-separated ionic polymer such as Nafion,\cite{jang2004, chou2005} where the backbone is hydrophobic and water molecules show a clear preference for the ionic domains. As a result, for the sulfonated polysulfones studied here an average of the dielectric constants of the polymer and water weighted based on their volume fractions provides a reasonable estimate of the dielectric constant ($\varepsilon$) that the ions experience. A similar conclusion was also reached by Kamcev \textit{et al.}.\cite{kamcev_effect_2017} In our analysis, the volume fractions are estimated using the mass of water and polymer present in a system and their respective densities. The dielectric constants employed in the weighted average are $\varepsilon_\text{polysulfone}=3.1$ and $\varepsilon_\text{water}=80.4$.\cite{noauthor_radel_2020} For the hydrated polymers simulated here, $\varepsilon$ computed in this way varies from 8.5 at $\lambda=3$ to 23 at $\lambda=14$. Correspondingly, the Bjerrum length ($l_\text{B}$) varies from 65~\AA~ at $\lambda=3$ to 24~\AA~ at $\lambda=14$, which are much larger than its value, about $7$~\AA, in dilute aqueous solutions of salts. The full set of $\varepsilon$ and $l_\text{B}$ values computed from the simulations is presented in Table \ref{TB2}.

\begin{table}[htb]
	\centering
	\caption{Parameters and variables obtained from MD simulations at different water contents ($\lambda$) for mBPS50 and dBPS25. }
		\begin{tabular}{|c|c|c|c|c|c|c|}
	\hline
		$\lambda$ & $\varepsilon$ & $l_\text{B}$ & $b$ (dBPS) & $b$ (mBPS) & $\xi$ (dBPS) & $\xi$ (mBPS) \\ 
		& & (\AA) & (\AA) & (\AA) & & \\ \hline
3      & 8.5               & 65.2 & 6.04     & 6.22     & 10.8           & 10.5           \\
4      & 10.2              & 54.7 & 6.10     & 6.21     & 9.0            & 8.8            \\
5      & 11.8              & 47.4 & 6.16     & 6.45     & 7.7            & 7.3            \\
6      & 13.3              & 42.0 & 6.32     & 6.50     & 6.6            & 6.5            \\
7      & 14.7              & 37.9 & 6.34     & 6.78     & 6.0            & 5.6            \\
8      & 16.1              & 34.6 & 6.67     & 6.76     & 5.2            & 5.1            \\
9      & 17.4              & 32.0 & 6.81     & 6.66     & 4.7            & 4.8            \\
10     & 18.7              & 29.8 & 6.79     & 6.99     & 4.4            & 4.3            \\
11     & 19.9              & 28.0 & 6.70     & 6.90     & 4.2            & 4.1            \\
12     & 21.1              & 26.4 & 6.78     & 6.88     & 3.9            & 3.8            \\
13     & 22.2              & 25.1 & 6.86     & 6.96     & 3.7            & 3.6            \\
14     & 23.3              & 23.9 & 7.02     & 7.24     & 3.4            & 3.3           \\ \hline
\end{tabular}
\label{TB2}
\end{table}

\subsection{Distribution of Ions}
\label{iondist}

In this section, we first compare the distribution of sulftonate ions in the disulfonated and monosulfonated polymers in terms of the sulfonate-sulfonate RDFs. Then we examine different methods for quantitatively determining the inter-ion distance ($b$) for the sulfonate ions.

\subsubsection{Comparison of Radial Distribution Functions}
\label{RDFcomp}

\begin{figure}[htb]
    \centering
    \includegraphics[width=.45\textwidth]{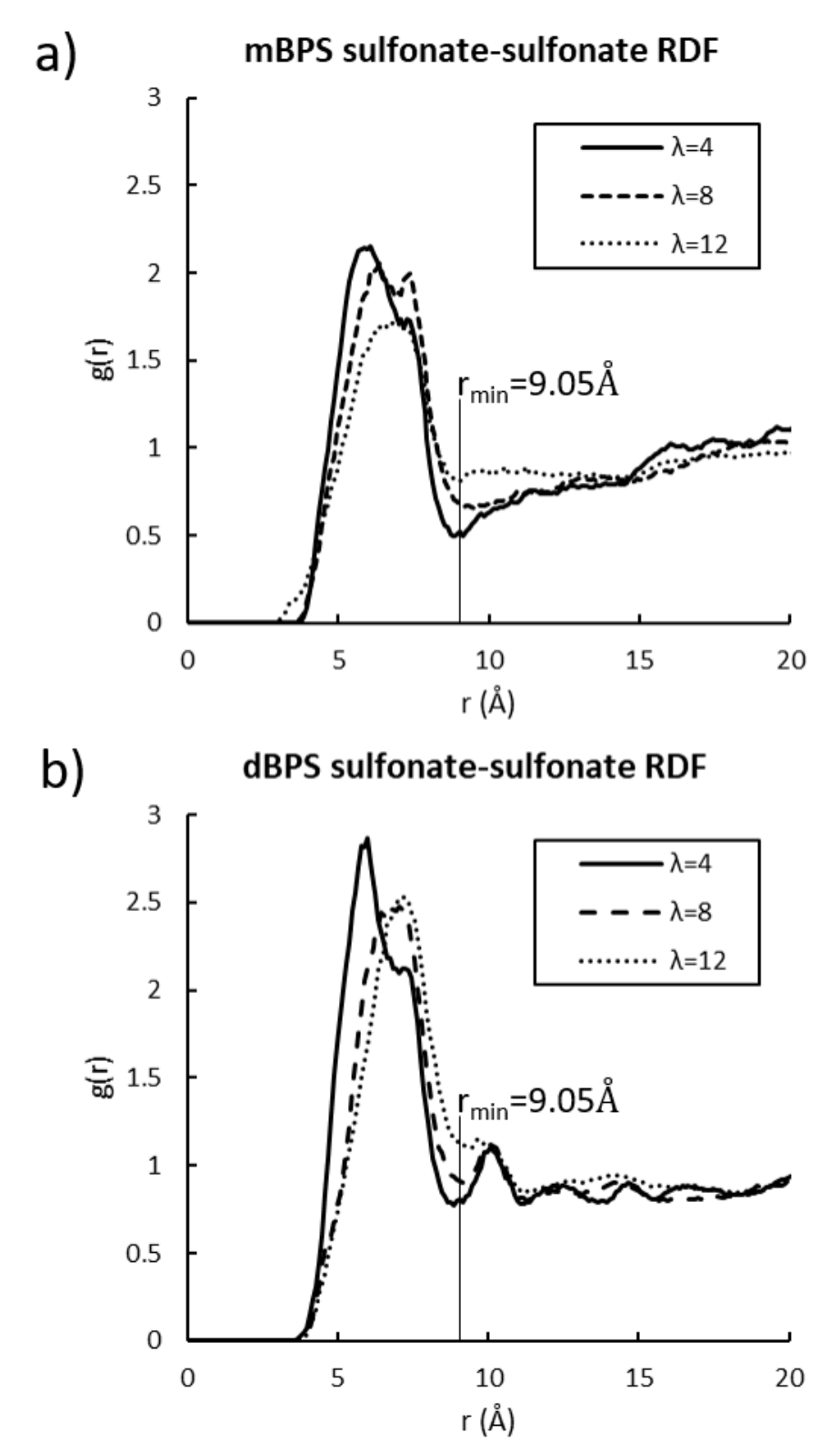}
    \caption{Sulfonate-sulfonate RDFs for the (a) mBPS50 and (b) dBPS25 tetramers at $\lambda=4$ (black trace), 8 (dark grey dashed trace), and 12 (light grey dotted trace).} 
    \label{fig:SO3SO3RDF}
\end{figure}

\par In order to probe the spatial distribution of the fixed ions, we plot the $\text{SO}_3^-$-$\text{SO}_3^-$ RDFs in Fig.~\ref{fig:SO3SO3RDF}. These RDFs can be read in a similar manner as a diffractogram in that more ordered structures create stronger and sharper peaks in their RDFs, whereas more disordered or amorphous structures yield broader features. The data previously presented in Figs.~\ref{fig:NaWcoord} and \ref{fig:SO3Wcoord} show that the monosulfonated and disulfonated polymers exhibit qualitatively similar hydration behavior. The $\text{SO}_3^-$-$\text{SO}_3^-$ RDFs also show certain similarity for the two polymers. However, there are discernible differences, as shown in Fig.~\ref{fig:SO3SO3RDF}. For example, the local ion density, which determines the height of the first peak of $g(r)$, is higher for the disulfonated polymer. The difference is quantified below in terms of the sulfonate-sulfonate coordination number.

\par The RDFs in Fig.~\ref{fig:SO3SO3RDF} show the subtle effect of water content on the ion distribution. At low water contents (e.g., $\lambda = 4$), the sulfonate-sulfonate RDF for the monosulfonated polymer in Fig.~\ref{fig:SO3SO3RDF}(a) shows a clear maximum near 6~\AA~and a minimum near 9~\AA, indicating certain aggregation behavior at short length scales for the sulfonate ions. Conversely, at higher water contents the features are less pronounced. The first peak has a reduced height, becomes broader, and is shifted to a larger separation, signifying a more random distribution of the sulfonate ions. This behavior is intuitive because in the monosulfonated case, the ions are located further away from each other on the polymer backbone. A chain can adopt various conformations allowing for a range of distances between the sulfonate ions. When the polymer swells as more water is added, the average separation between the sulfonate ions on the same chain or different chains also increases, leading to the shift of the peak location of the corresponding $g(r)$.

\par In contrast, the sulfonate-sulfonate RDF for the disulfonated polymer (Fig.\ref{fig:SO3SO3RDF}(b)) retains its sharp primary peak even up to high water contents, although its location also shifts to a somewhat larger separation as the water content is increased. This behavior originates from the constraint on the inter-sulfonate distance in the disulfonated polymer, where the two sulfonate ions on each chain are located on the same sulfone monomer (see Fig.~\ref{fig:structure_MD}). The small amount of conformational variability of the chain segment between the two sulfonate ions allows their separation to slightly increase when the hydration level is increased. However, the inherently paired nature of the sulfonate ions on the same chain does not allow them to significantly move away from each other. As a result, the primary peak of the sulfonate-sulfonate RDF for the disulfonated polymer almost retains its height event at high levels of hydration, even though the overall swelling with more water added makes the peak location to shift toward slightly larger separations.

\par The difference in the ion distribution between the monosulfonated and disulfonated polymers is especially apparent at high water contents (e.g., compare the light grey dotted traces in Figs.~\ref{fig:SO3SO3RDF}(a) and (b) for $\lambda = 12$), where the disulfonated polymer maintains a high ion density near 7.5~\AA. This distance is consistent with the inter-sulfonate distance of a disulfonated monomer near its energy-minimized conformation from the quantum mechanical calculations. In contrast, the $\text{SO}_3^-$-$\text{SO}_3^-$ RDF for the monosulfonated polymer at $\lambda=12$ is much flatter, indicating a more uniform distribution of ions throughout the polymer. At lower water contents, the inter-sulfonate distance is similar for both monosulfonated and disulfonated polymers, so the difference in their sulfonate-sulfonate RDFs is less apparent. The fact that the two polymers show more different RDFs for the sulfonate ions at high water uptakes is consistent with the experimental observation that the bulk properties of the monosulfonated and disulfonated polymers are more similar at low ion contents and low water uptakes, but begin to diverge when the ion content and water uptake are increased.

\begin{figure}[htb]
   \centering
   \includegraphics[width=.45\textwidth]{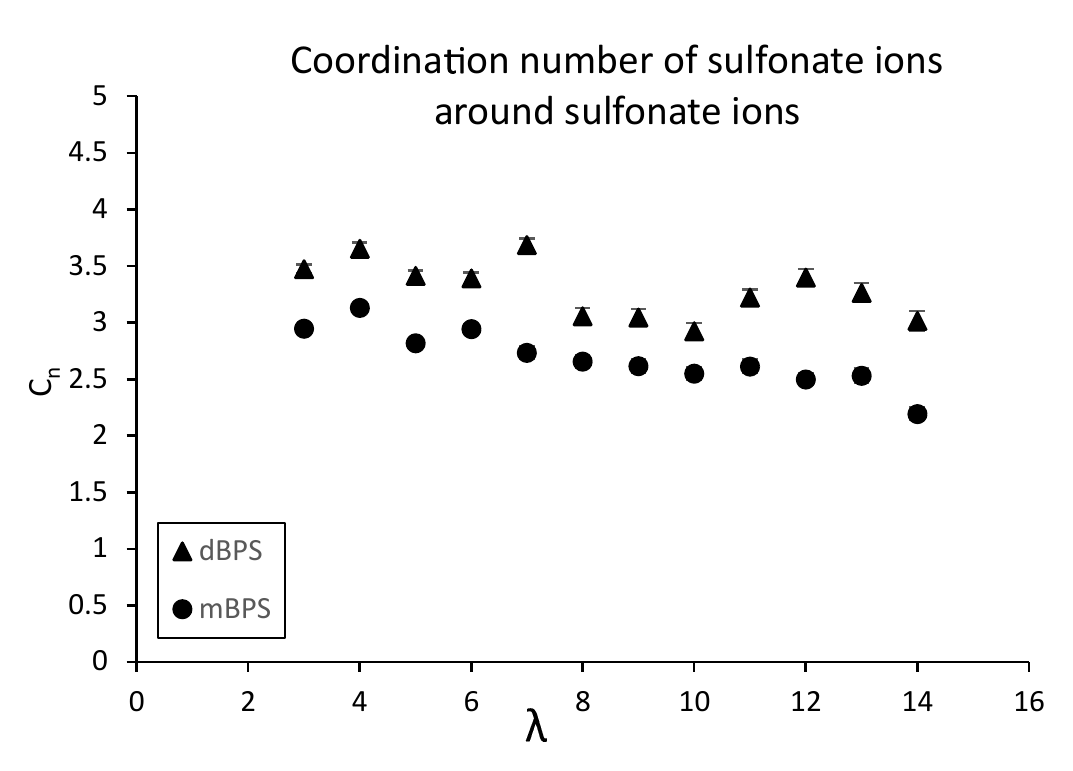}
   \caption{Sulfonate-sulfonate coordination number ($C_n$) vs. water uptake ($\lambda$) for the dBPS25 (triangles) and mBPS50 (circles) tetramers. Cutoff radius ($r_\text{min}$) is 9.05 \AA.} 
   \label{fig:SO3SO3Coord}
\end{figure}

\par The difference between the fixed-ion RDFs of the monosulfonated and disulfonated polymers is also reflected by the coordination number of sulfonate ions around other sulfonate ions (Fig. \ref{fig:SO3SO3Coord}). The location of the first minimum in $g(r)$ shows some variations, so we use a constant cutoff distance of 9.05~\AA~for comparison purposes. The results are shown in Fig.~\ref{fig:SO3SO3Coord}. For both polymers, the sulfonate coordination number slightly decreases as the water content increases. This result reflects the fact that the separation between sulfonate groups becomes larger when more water is added (see Fig.~\ref{fig:SO3Wcoord}). Overall, the sulfonate coordination number for the disulfonated polymer is higher than that for the monosulfonated polymer and the difference becomes more pronounced at higher water contents. This is a reflection of the trend in the RDFs discussed in the previous paragraph.

\subsubsection{Calculation of Inter-Ion Distance}
\label{calcb}

\par
\begin{figure*}[htb]
    \centering
    \includegraphics[width=\textwidth]{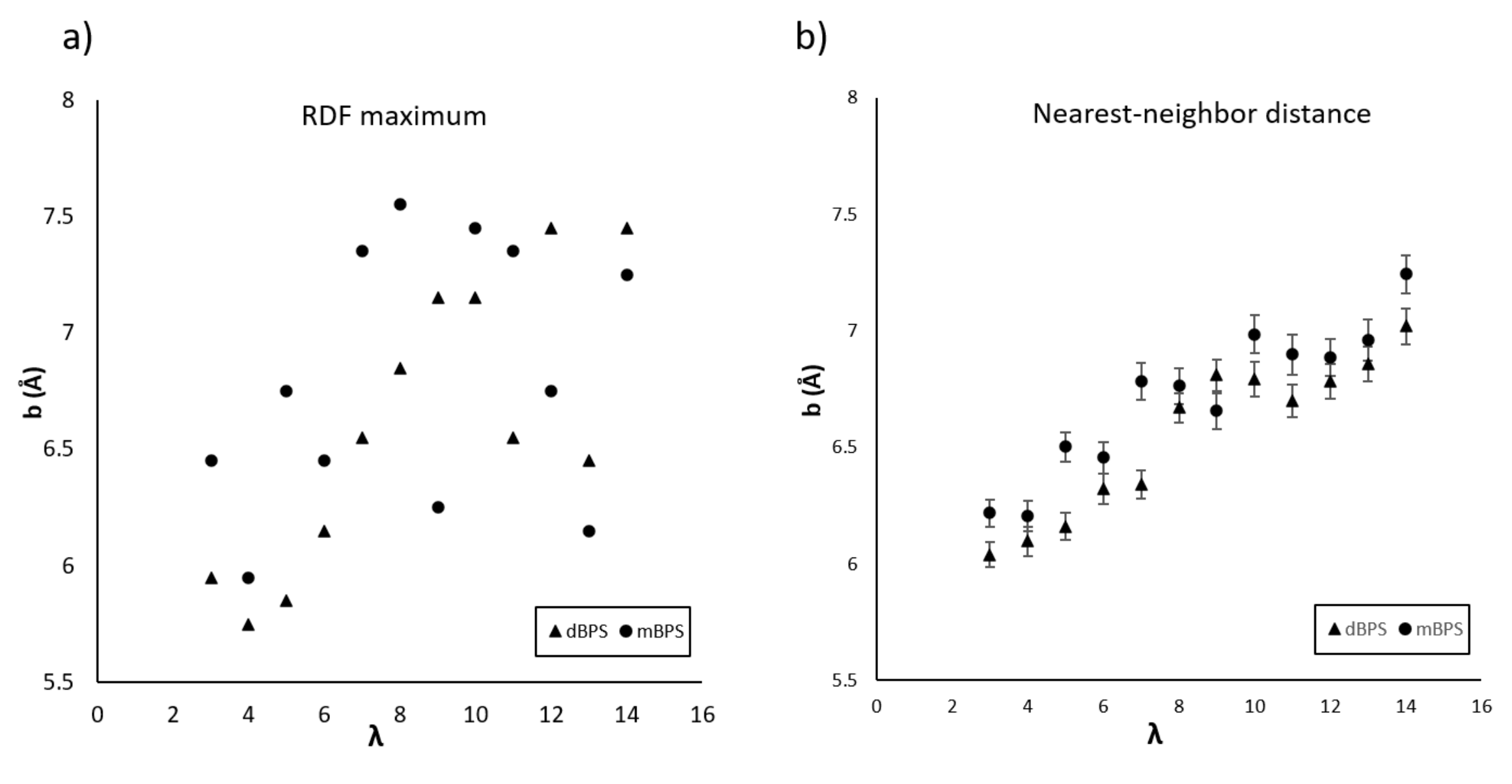}
    \caption{A comparison of two methods for calculating the average inter-sulfonate distance ($b$), as a function of water uptake ($\lambda$), on the basis of the sulfonate-sulfonate RDF. The value of $b$ is estimated using a) the location of the RDF maximum and b) the average nearest-neighbor distance, which is the radial distance at which $C_n$=1.} 
    \label{fig:compareb}
\end{figure*}

 Here we discuss how to use the sulfonate-sulfonate RDFs to estimate the average distance between the fixed ions, $b$, and use it to compute the Manning parameter, $\xi$. From Fig.~\ref{fig:SO3SO3RDF}, it is clear that assigning a single characteristic inter-ion distance to a complex three-dimensional distribution of ions is not straightforward. We compare two different methods of computing $b$ from the sulfonate-sulfonate RDFs:
\begin{enumerate}
    \item The location of the maximum in the sulfonate-sulfonate RDF.
    \item The average nearest-neighbor distance (i.e., the value of $r_\text{min}$ in Eq.~(\ref{eq:coord_number}) at which $C_n$=1).
\end{enumerate}
The resulting values of $b$ are plotted in Fig.~\ref{fig:compareb}. The location of the RDF maximum (Fig.~\ref{fig:compareb}(a)) provides a poor estimate of $b$, especially in the monosulfonated case where the broad multimodal shape of the RDF makes the location of the maximum particularly unreliable.

\par A more reliable estimate of $b$ is provided by the average nearest-neighbor distance (Fig.~\ref{fig:compareb}(b)), which is calculated by finding the separation at which the sulfonate-sulfonate coordination number ($C_n$) reaches 1. The nearest-neighbor method is insensitive to the location of the extrema of RDFs, which improves the reliability of the resulting estimate. This estimate yields a consistently larger value of the average $\text{SO}_3^-$-$\text{SO}_3^-$ distance for the monosulfonated polymer than the disulfonated one. The difference is expected based on the fact that the sulfonate groups are more widely separated along the polymer backbone in the monosulfonated case. Therefore, we use the values of $b$ from the nearest-neighbor method in subsequent analyses and reproduce them in Table \ref{TB2} for reference.

\begin{figure}[htb]
    \centering
    \includegraphics[width=.45\textwidth]{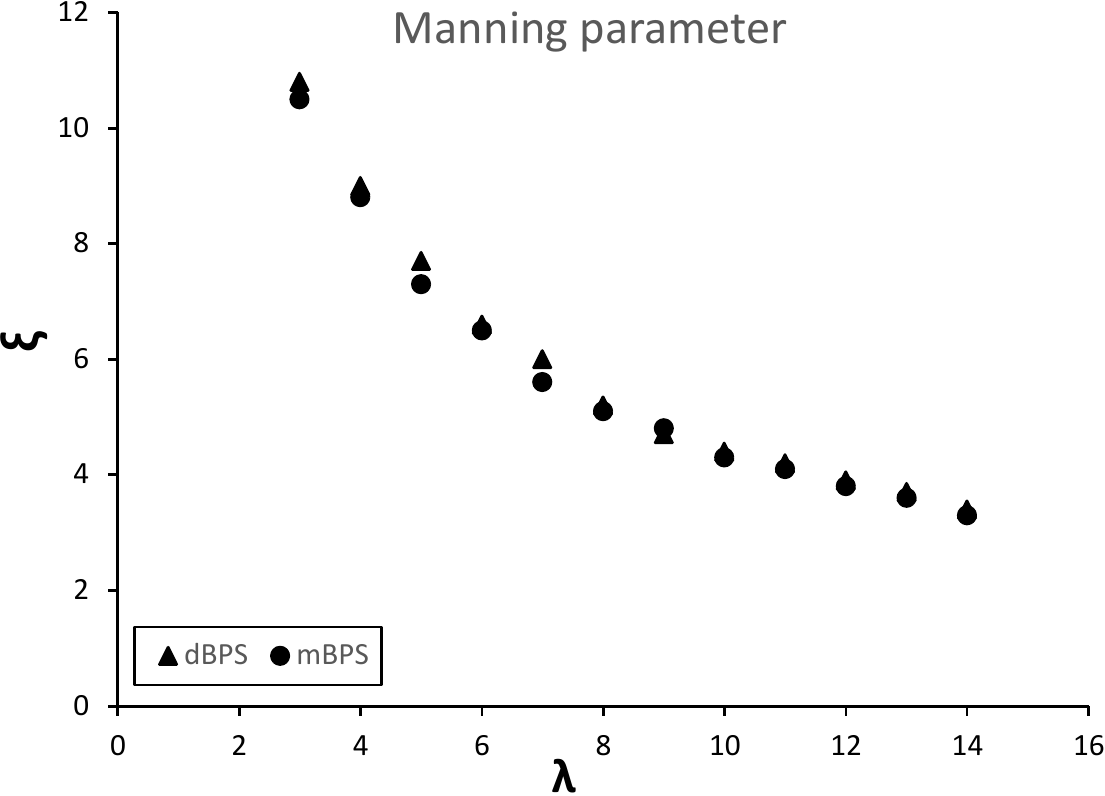}
    \caption{Manning parameter ($\xi$) based on the value of $b$ in Fig.~\ref{fig:compareb}b vs. water uptake ($\lambda$).} 
    \label{fig:Manning}
\end{figure}

\par In general, the average separation between the sulfonate ions is much smaller than the Bjerrum length (see Table \ref{TB2}). Since the Manning parameter is defined as $\xi \equiv l_\text{B}/b$, the resulting value of $\xi$ is significantly larger than the critical value for monovalent ions ($\xi_\text{c}=1$), as shown in Fig.~\ref{fig:Manning}. This means that a large portion of the counterions are condensed to the fixed ions. In the monosulfonated case, a slightly larger value for the inter-sulfonate distance results in a slightly smaller Manning parameter, giving the following relation
\begin{equation}
    \xi_\text{dBPS}>\xi_\text{mBPS}\gg\xi_\text{c}=1~.
\end{equation}
Since the fraction of ions that is condensed is $1-\xi_\text{c}/\xi$ (Eq. \ref{eq:condensedfrac}), the above relation indicates that the disulfonated polymer exhibits stronger counterion condensation. Condensed ions do not contribute to Donnan exclusion, so the higher the value of $\xi$, the weaker the coion exclusion. In the monosulfonated polymer the more even spatial distribution of the fixed ions results in a smaller value of $\xi$, which implies fewer condensed counterions and enhanced Donnan exclusion of coions. We will continue to explore this difference in the next section.

\begin{figure}[htb]
    \centering
    \includegraphics[width=.45\textwidth]{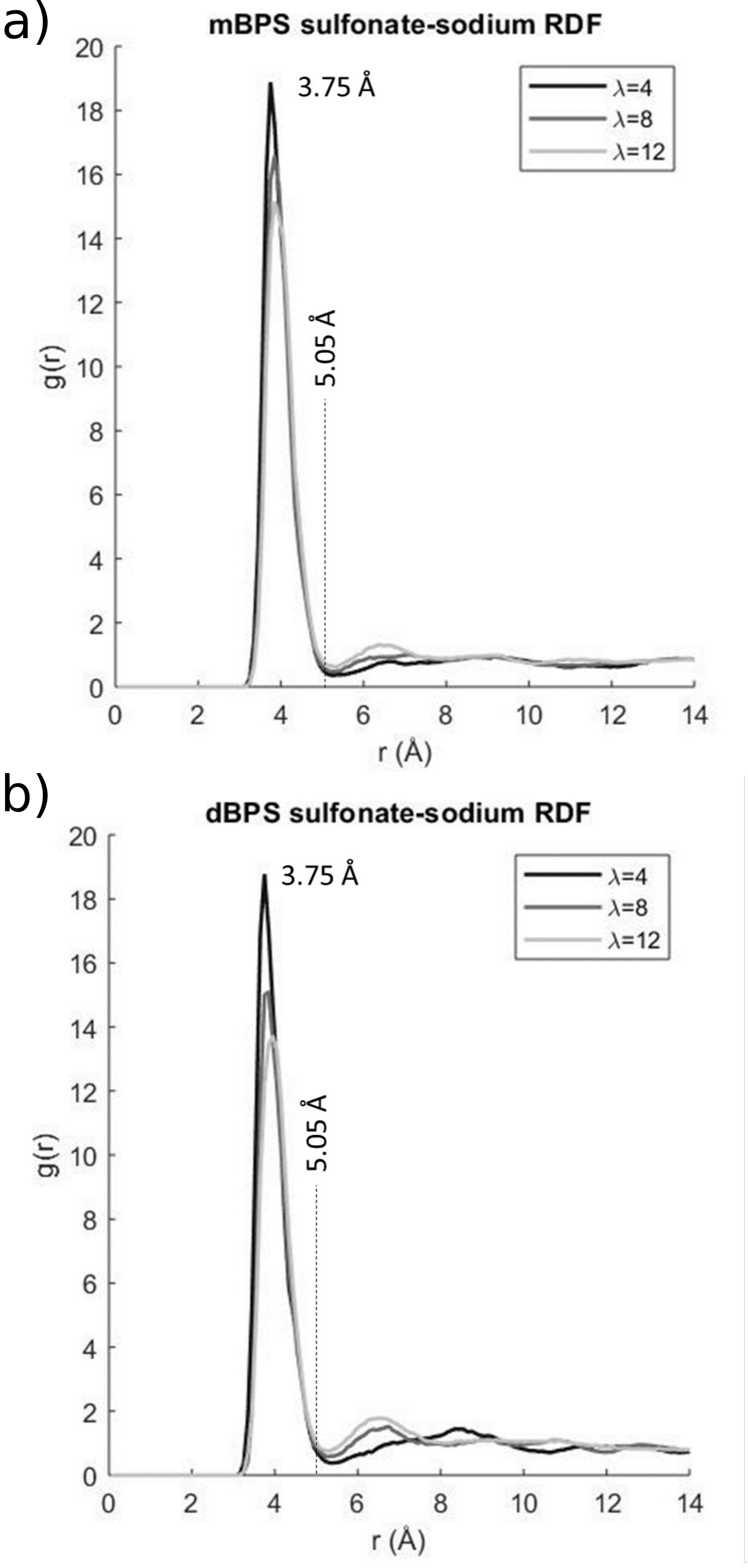}
    \caption{Sulfonate-sodium RDFs for the (a) mBPS50 and (b) dBPS25 tetramers at $\lambda=4$ (black trace), $8$ (dark grey trace), and $12$ (light grey trace).} 
    \label{fig:SO3NaRDF}
\end{figure}

\subsection{Counterion Condensation (Anion-Cation Pairing)}
\label{pairing}

\par
\begin{figure}[htb]
    \centering
    \includegraphics[width=.45\textwidth]{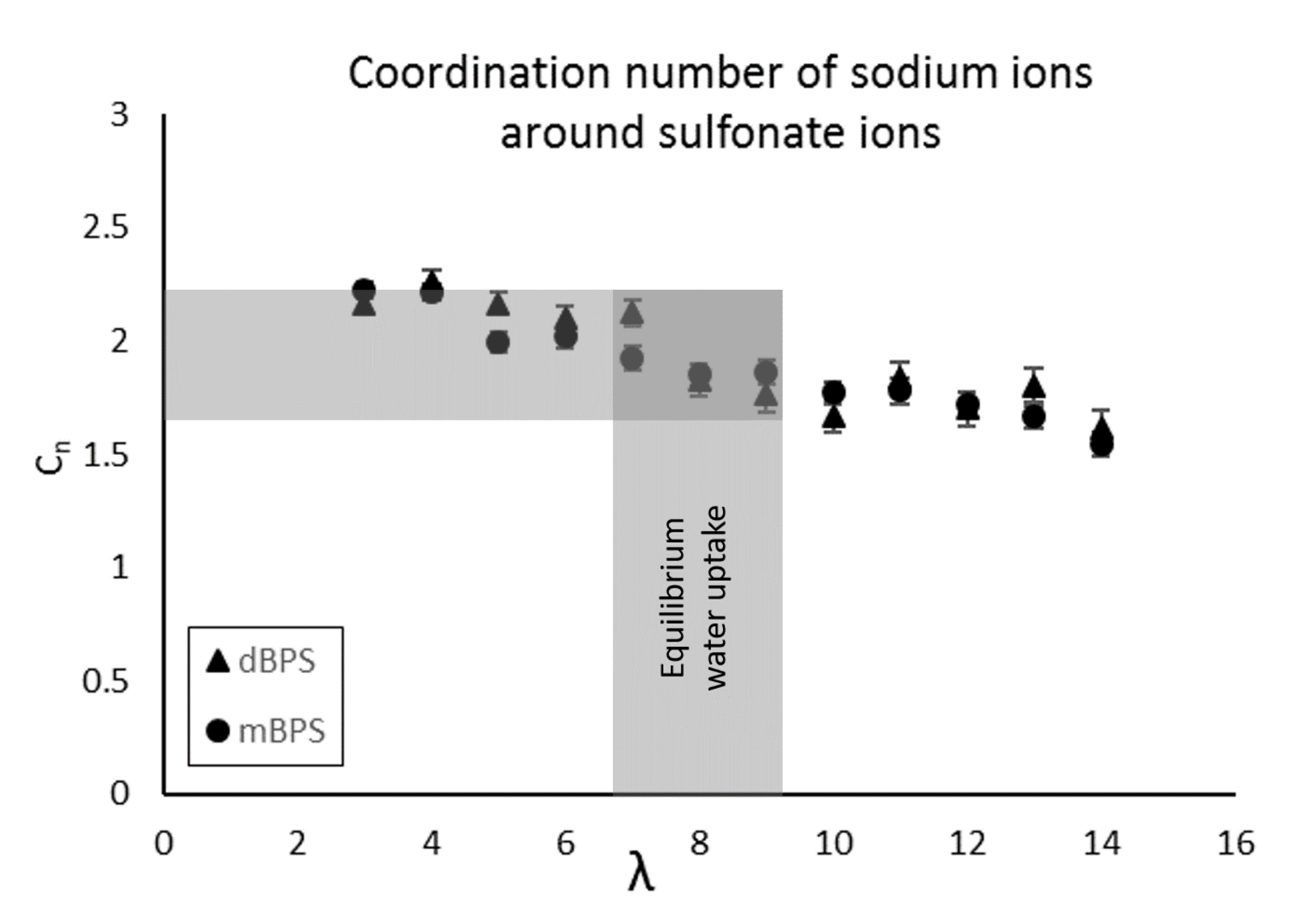}
    \caption{Coordination number of the sodium counterions around the sulfonate groups for the dBPS25 (triangles) and mBPS50 (circles) tetramers. The cutoff radius ($r_\text{min}$) is 5.05~\AA.} 
    \label{fig:SO3NaCoord}
\end{figure}

 To further investigate the counterion condensation phenomena within the sulfonated polysulfones, we examine in this section the distribution of the mobile sodium counterions around the fixed sulfonate ions, which is quantified using the sulfonate-sodium RDFs shown in Fig.~\ref{fig:SO3NaRDF}. The single sharp peak around $r=4$~\AA~indicates that the fixed ions and counterions exhibit direct contact, rather than water-mediated, interaction. In traditional polyelectrolyte systems, the condensed counterions are considered to be ``paired" with the fixed ions. However, by integrating the $\text{SO}_3^-$-$\text{Na}^+$ RDF to a cutoff radius ($r_\text{min}$ in Eq.~\ref{eq:coord_number}) of 5.05 \AA, we find that the coordination number of the sodium ions around the sulfonate ions is approximately 2 at the equilibrium water uptake (Fig.~\ref{fig:SO3NaCoord}). This indicates that many of the fixed ions and counterions are not simply paired but, on average, a cation is shared between two anions and an anion is shared between two cations. The formation of such small ionic condensates was previously indicated by the experimental investigation of Marino \textit{et al.}.\cite{marino_hydroxide_2014} It was also noted in the simulations of a disulfonated diphenyl sulfone by Wohlfarth \textit{et al.}, who referred to the phenomenon as ``triple ions''.\cite{wohlfarth_proton_2015}

\par
\begin{figure*}[htb]
    \centering
    \includegraphics[width=\textwidth]{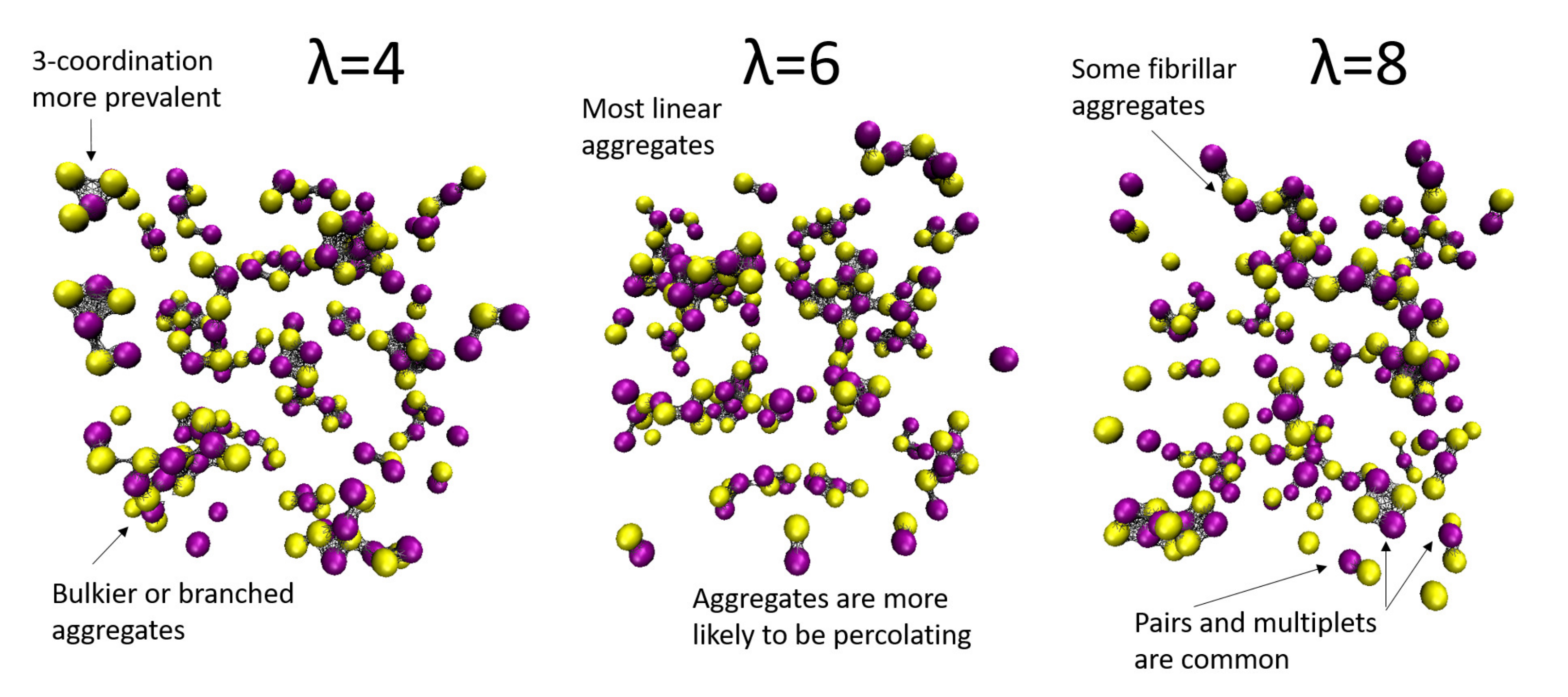}
    \caption{Visualization of ionic aggregates in the MD simulations of the mBPS50 tetramers at $\lambda=4$, 6, and 8. Color code: sulfur atoms in the sulfonate groups (yellow); sodium counterions (purple). Grey webs are added as a guide to the eye to highlight the adjacency of ions within the 5~\AA~radius, the location of the first minimum of the RDFs in Fig.~\ref{fig:SO3NaRDF}. The polymer backbone chains and water molecules are not visualized to improve clarity.}
    \label{fig:chainslambda}
\end{figure*}

\begin{figure}[htb]
    \centering   
    \includegraphics[width=.45\textwidth]{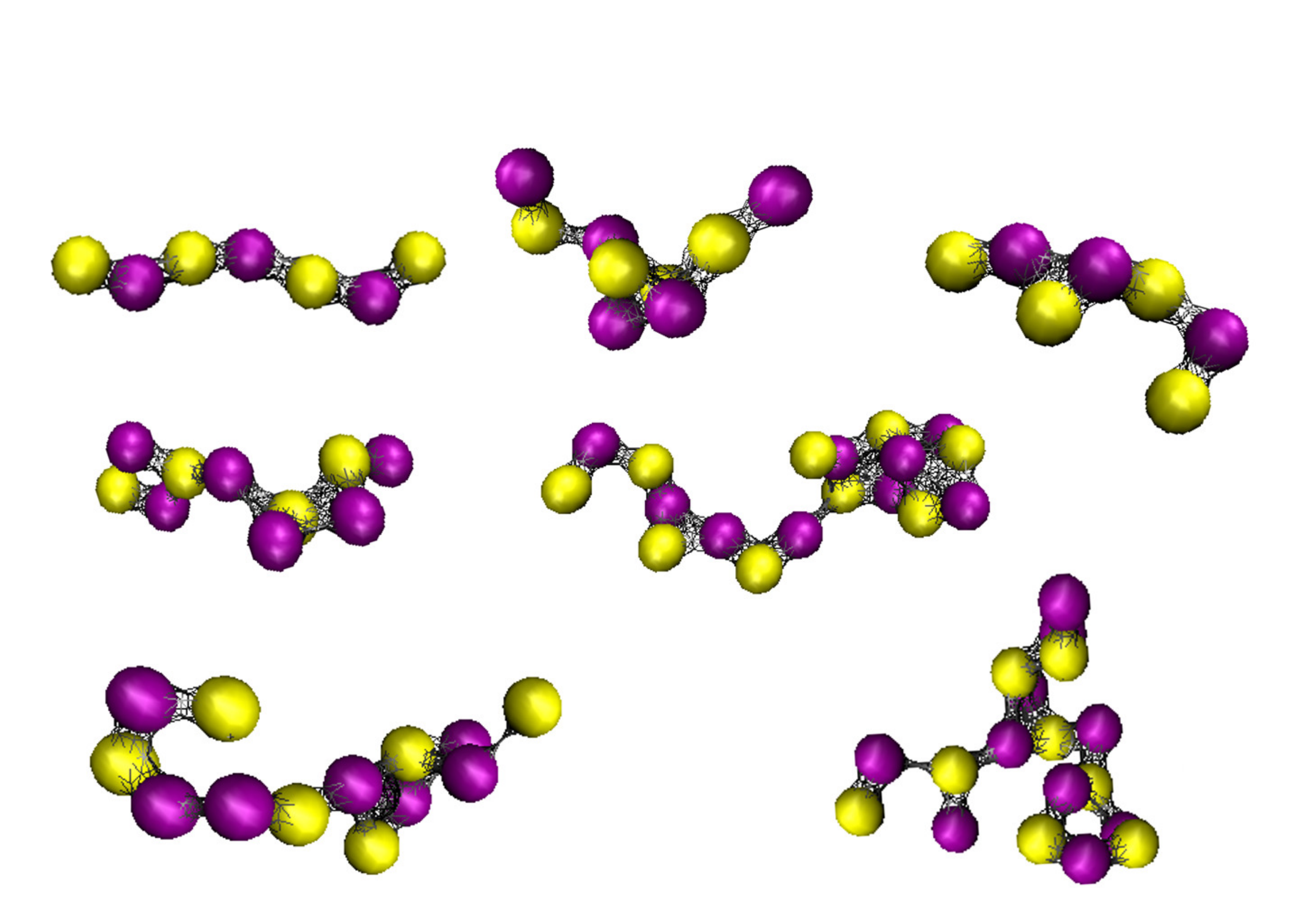}
    \caption{Examples of some chain-like aggregates observed in the MD simulations of the mBPS50 tetramers at $\lambda$=8. Color code: sulfur atoms in the sulfonate groups (yellow); sodium counterions (purple). Grey webs are added as a guide to the eye to highlight the adjacency of ions.}
    \label{fig:bigchain}
\end{figure}

 The fact that the coordination number of $\text{Na}^+$ around $\text{SO}_3^-$ is near 2 implies that rather than forming isolated pairs, the ions can form chain-like aggregates of alternating charges within the hydrated polymer matrix. Such fibrillar ionic aggregates are visible in our MD simulations at low water contents ($\lambda \lesssim 7$), and several snapshots illustrating this effect are shown in Figs.~\ref{fig:chainslambda} and \ref{fig:bigchain}. Ion sharing and chain-like structures are particularly clear in Fig.~\ref{fig:bigchain}. Similar structures of ions have also been observed in polyethylene oxide-based ionomers and concentrated salt solutions.\cite{lin_cation_2012,choi_graph_2018}

\par Figure \ref{fig:SO3NaCoord} further shows that the aggregation behavior of the ions depends on the water content in the polymer. As low water contents with $\lambda \lesssim 7$, $C_n \gtrsim 2$, which is representative of the membrane at equilibrium with a humid ambient environment. In this case the ions form comparatively large dense aggregates in the hydrated polymers. Near the equilibrium water uptake with $\lambda \simeq 7$, $C_n\approx2$ and the ionic chains begin to break apart but some short fibrillar aggregates remain. At high water contents with $\lambda \gtrsim 7$, $C_n<2$ and the ions are more dispersed and only form small aggregates, on the order of two or three ions per cluster. Values of $\lambda$ higher than $\lambda_\text{eq}$ are not representative of experimentally obtainable polymer-water systems. However, the simulations allow us to probe the region of high water uptakes and reveal that the ions are less aggregated when more water is added. 

\par
\begin{figure}[htb]
    \centering   
    \includegraphics[width=.45\textwidth]{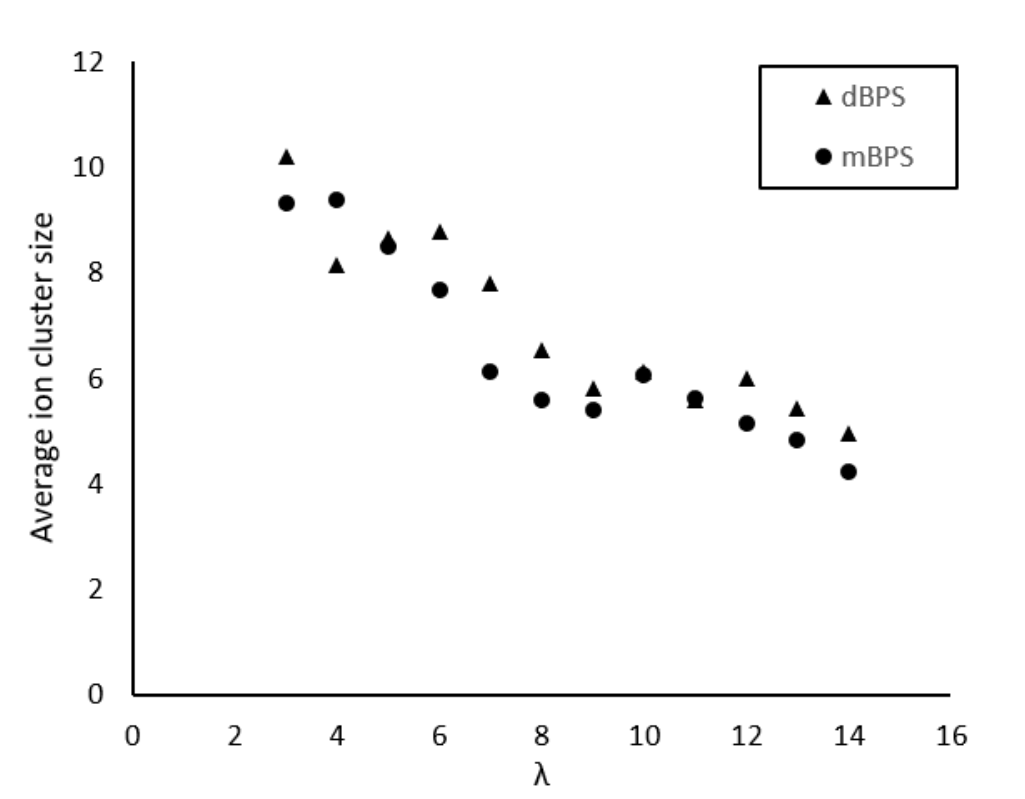}
    \caption{Average size of ion clusters vs. water uptake ($\lambda$).}
    \label{fig:cluster}
\end{figure}

 In order to further quantify the extent of ion clustering in the simulations, we conduct a distance-based cluster analysis to estimate the average cluster size.\cite{abbott_nanoscale_2017} A $\text{Na}^+$ ion and a $\text{SO}_3^-$ ion are defined as clustered if they are within the 5.05~\AA~cutoff distance, the location of the first minimum of the $\text{SO}_3^-$-$\text{Na}^+$ RDF in Fig.~\ref{fig:SO3NaRDF}. For each snapshot, all the clusters are identified using this distance criterion of clustering. The cluster size, in terms of the number of sodium and sulfonate ions in a cluster, is averaged over all the clusters and the 1000 snapshots collected in the last 1 ns of our MD simulations of each system. The result is plotted in Fig.~\ref{fig:cluster} against the water uptake for both disulfonated and monosulfonated polymers. As the water uptake is increased, the average size of ion clusters decreases, consistent with the visualization in Fig.~\ref{fig:chainslambda} and the trend of the inter-ion distance shown in Fig.~\ref{fig:compareb}b. Although somewhat noisy, the data also show that in general, the average cluster size is slightly larger for the disulfonated polymers than the monosulfonated ones at the same level of hydration.

\par The existence of ionic chains and networks, in which the fixed ions are bridged by the mobile ions, at water contents up to the equilibrium water uptake may help explain the experimental observation that ion transport in a membrane based on aromatic polymers such as polysulfones can be high despite the glassy nature of the polymers.\cite{Wang2001, daryaei2017-1, daryaei2017-2, daryaei2017Diss, choudhury2019} The common belief is that if cations and anions are paired, then there must be a mechanism for transferring mobile ions from one fixed ion to another (e.g., by hopping or vehicular mechanisms). However, if the lowest-energy state actually involves chains of mobile ion shared between fixed ions, as indicated by our simulations and others,\cite{wohlfarth_proton_2015} then the distinction between ion ``transport'' and ion ``conduction'' is somewhat blurred. This type of ion conduction was recently demonstrated by Dokko \textit{et al.} in concentrated liquid electrolytes.\cite{dokko_direct_2018} Based on our simulations, we hypothesize that a similar mechanism can enhance ion transport in glassy ionic polymers.

\par Using the values of $\varepsilon$ and $b$ determined previously, we have computed the fraction of counterions that is condensed to the fixed ions using the Manning theory. This fraction can also be quantified directly by analyzing the configurations generated by the MD simulations, such as those in Fig.~\ref{fig:chainslambda}. First, a criterion has to be established to separate the counterions into the condensed and uncondensed groups. To this end, we define a $\text{Na}^+$ counterion as fully condensed if it is in the middle of an ionic chain or aggregate (see Fig.~\ref{fig:bigchain}). That is, the fully condensed counterions are at least doubly coordinated by the fixed ions. Furthermore, a $\text{Na}^+$ counterion at the ends of an ionic chain or the terminals of an aggregate, which is singly coordinated, is half condensed since it shares its neighboring $\text{SO}_3^-$ ion with another $\text{Na}^+$ counterion. The number of uncondensed or ``free" sodium counterions, $N_\text{f}$, is then given by
\begin{equation}
    N_\text{f}=N_0+0.5N_1~,
    \label{eq:countcond}
\end{equation}
where $N_0$ is the number of sodium counterions not associated to any fixed ions and $N_1$ is the number of sodium counterions coordinated with just one sulfonate ion. Normalizing $N_\text{f}$ by the total number of sodium counterions in the system, we obtain the fraction of sodium counterions that is uncondensed.

 \begin{figure}[htb]
    \centering   
    \includegraphics[width=.45\textwidth]{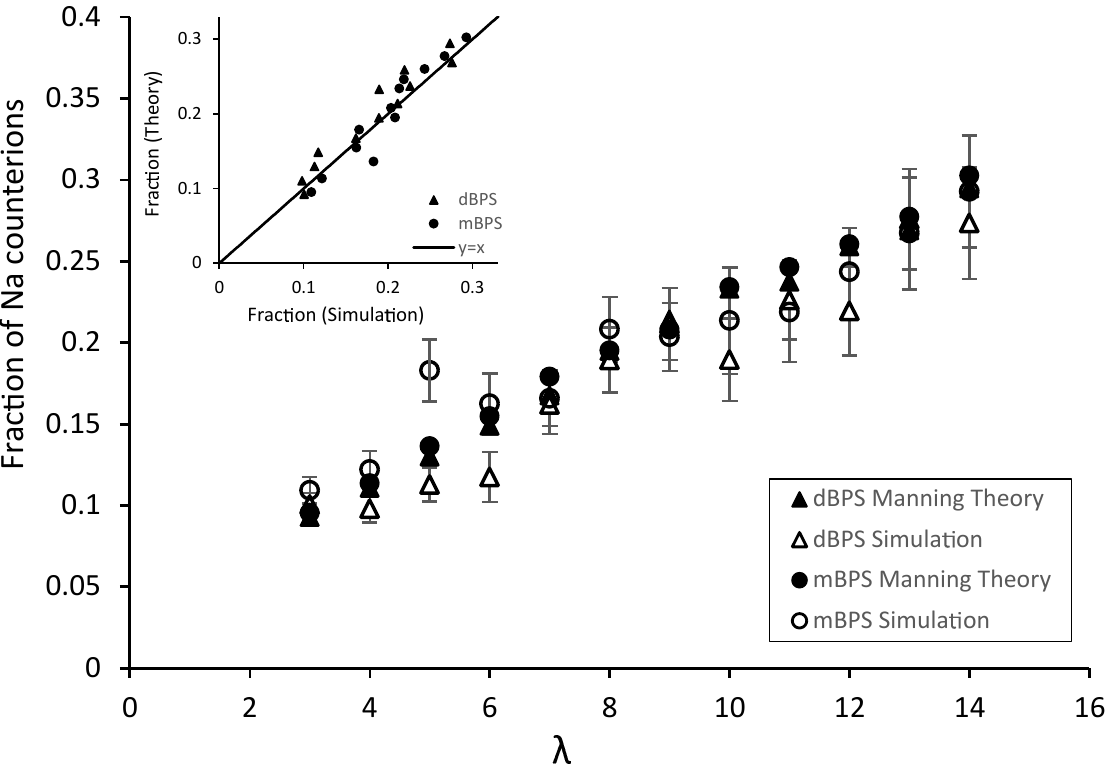}
    \caption{Fraction of sodium counterions that is uncondensed is plotted against water content ($\lambda$). Filled symbols represent the prediction of Manning's counterion condensation theory, and open symbols represent the results based on uncondensed ions identified directly from the MD simulations. Error bars of the open symbols represent the standard deviation of the results from the 1,000 snapshots analyzed. The inset shows the direct comparison of the theory and simulations regarding this fraction and a good agreement is found.}
    \label{fig:condensed}
\end{figure}

\par The fraction of $\text{Na}^+$ counterions that is uncondensed determined directly from the MD snapshots generally agrees well with the prediction of the Manning theory based on the values of $\varepsilon$ and $b$ obtained previously, as shown in Fig.~\ref{fig:condensed}. This is especially the case for the monosulfonated polymer. The number of uncondensed ions calculated directly from the simulations for the disulfonated polymer appears to have somewhat more variability. This variability may be due to the more complex geometry of the ionic aggregates in the disulfonated case, which are not well captured by the theory behind Eq.~\ref{eq:countcond}. Also, some of the variability likely originates from the fact that the number of tetramers included in our MD simulations, and the resulting number of sulfonate ions, is still small because of the limited availability of computational resources. Larger simulations would have improved statistics and may yield more precise results on the fraction of counterions that is uncondensed.

 \begin{figure}[htb]
    \centering   
    \includegraphics[width=.45\textwidth]{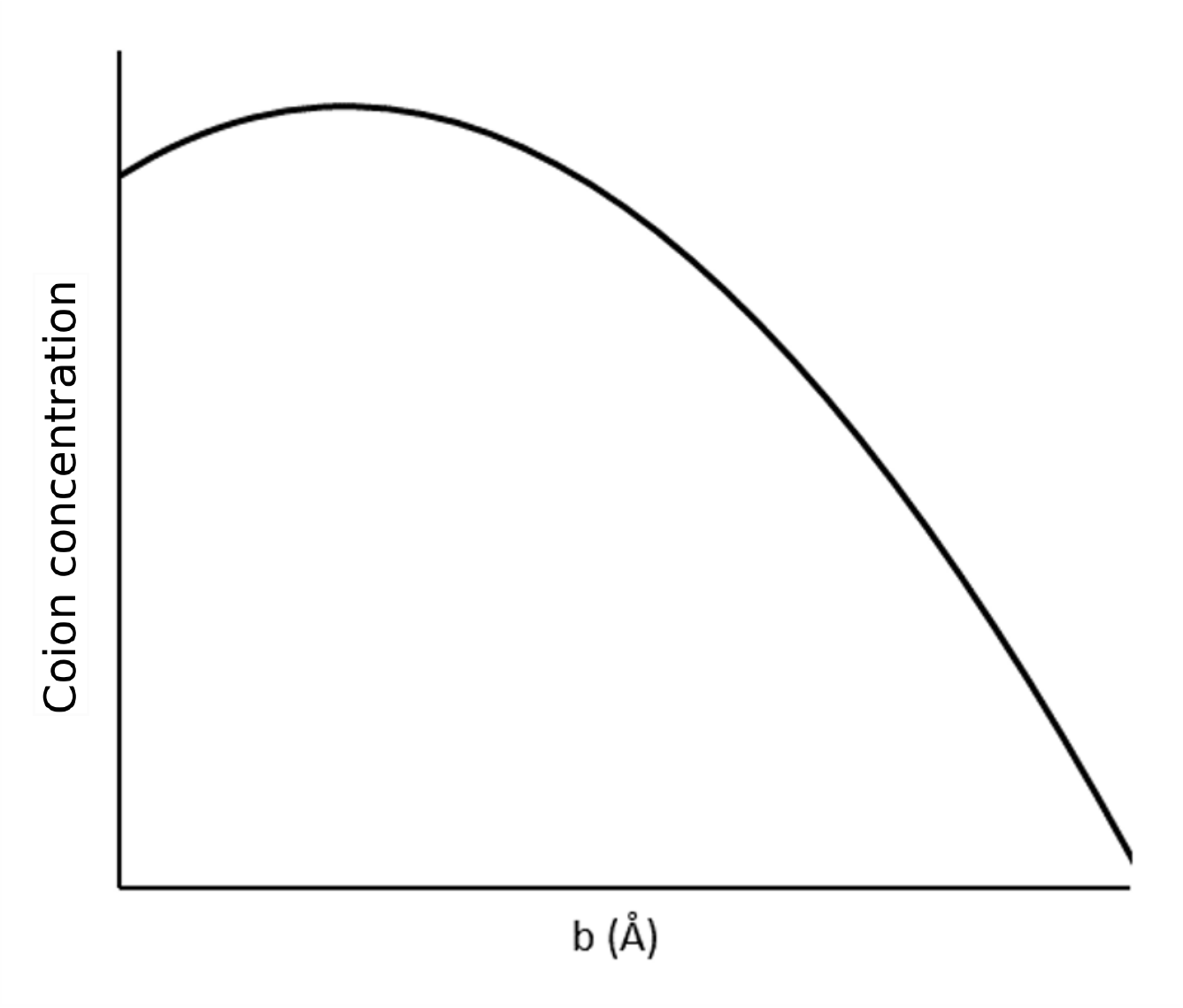}
    \caption{Coion concentration ($C_-^\text{m}$) in a sulfonated polysulfone membrane placed in a concentrated salt solution vs. the inter-sulfonate distance ($b$). For experimentally relevant values of $b$, they are roughly inversely related as $C_-^\text{m} \propto 1/b$.}
    \label{fig:Cvsb}
\end{figure}
    
%    \left[\frac{\gamma_{\frac{+}{}}(C_{\text{SO}_3}^\text{m})^2 e^{\frac{1}{\frac{2 l_\text{B}}{b} + 1}}}{b/l_\text{B} + 2} + \frac{(C_{\text{SO}_3}^\text{m})^2}{4})\right]^{1/2} \nonumber   & - \frac{C_{\text{SO}_3}^\text{m}}{2}~,

\par For a sulfonated polysulfone membrane in a concentrated NaCl solution, the coion concentration in the membrane is\cite{kamcev_partitioning_2016} 
\begin{align}
    \lim_{C_\text{NaCl}^\text{s}\rightarrow C_{\text{SO}_3}^\text{m}}C_-^\text{m} \approx C_{\text{SO}_3}^\text{m} \Bigg\{ \left[ 2\gamma^2 \frac{e^{b/(b+2l_\text{B})}}{b/l_\text{B}+1}+\frac{1}{4}\right]^{1/2} -\frac{1}{2} \Bigg\}
\label{eq:C}
\end{align}
where $\gamma$ is the mean activity coefficient of the salt in the adjacent aqueous solution and $C_{\text{SO}_3}^\text{m}$ is the concentration of the fixed sulfonate groups within the membrane. The relationship between $C_-^\text{m}$ and $b$ reflects how the inter-ion spacing impacts the ion exclusion property of the membrane. A representative curve showing the relationship between $C_-^\text{m}$ and $b$ is shown in Fig.~\ref{fig:Cvsb}. For the systems that we are interested in, the maximum in $C_-^\text{m}$ occurs at a value of $b$ close to the hard sphere radius of the ions. Since for both monosulfonate and disulfonated polymers, the value of $b$ is larger than the ions' hard sphere radius, $C_-^\text{m}$ and $b$ are approximately inversely related ($C_-^\text{m} \propto 1/b$). As a result, a membrane with a larger value of $b$ exhibits better coion exclusion. Since $b$ is larger for the monosulfonated polysufones, a membrane based on them is expected to exhibit stronger coion exclusion than that based on the disulfonated polymers at a similar ion content.

\section{Conclusions}

\par We have performed large-scale MD simulations to probe the hydration, ion distribution (including the average spacing between the fixed ions) and cation-anion interaction in two sulfonated polysulfones with the same ion content, but different ion distributions along the backbones. We apply Manning's counterion condensation theory,\cite{manning_limiting_1969} which was initially proposed for polyelectrolyte solutions, to analyze the simulations of these glassy ionic polymers. Our results shed light on the counterion condensation phenomena in dense ionic polymers and yield a deeper understanding of the molecular origin of the subtle differences in the bulk properties of the two polymers being investigated.

\par The simulations indicate that at the equilibrium water content, the ions have a similar level of hydration as they would in their saturated aqueous solutions. The simulations also indicate that water molecules do not have a strong preference for the ions and at least 25\% of the water molecules in the polymer matrix are associated with the polar groups on the polymer backbone. The simulations thus indicate that there is not a distinct ionic phase as there would be in microscale phase-separated ionic polymers. For the ionic polymers studied here, the dielectric constant ($\varepsilon$) of the environment that the ions experience can be well approximated by a volume-weighted average of the dielectric constants of the polymer and water with the volume fractions computed from their respective masses and densities.

\par Based on the simulation results, the spacing of the sulfonate ions along the polysulfone backbone impacts the spatial distribution of the sulfonate ions inside the hydrated polymer. At low water contents up to the equilibrium water uptake, the shape of the RDF for the monosulfonated polymer is similar to that of the disulfonated polymer. However, as more water is added and the polymer swells, the ions on the same chain in the monosulfonated polymer can move away from each other, while those in the disulfonated polymer cannot. This difference is responsible for the noticeably different distributions of the sulfonated ions in the two polymers at higher water contents. The inter-ion distance, $b$, between the sulfonate groups, can be reliably computed from the sulfonate-sulfonate RDF as the separation at which the sulfonate-sulfonate coordination number reaches 1. Using this method, we find that the inter-sulfonate ion distance is slightly larger for the monosulfonated polymer over the wide range of water contents simulated here.

\par Our simulations reveal that at low water uptakes (below the equilibrium water uptake for the polymers under investigation), the sulfonate ions and sodium counterions form ionic chains and networks via direct contact, rather than water-mediated interactions. At low water contents, such ionic networks are hypothesized to facilitate ion transport without the need for polymer-backbone motion. This mechanism of ion transport along fibrillar aggregates appears to be specific to glassy polymers and other concentrated electrolytes with low water contents. This understanding may inform development of ionic polymer membranes for applications involving ion transport and bridge the gap between ion-conducting glasses and compliant ionic polymer materials.

\par With $\varepsilon$ and $b$ obtained from simulations, we can estimate the fraction of counterions that is either condensed or uncondensed using the Manning theory. The prediction is compared to the simulation result by examining the ion-ion RDFs and counting the sodium counterions that are either free or coordinated with only 1 sulfonate group. A good agreement is found between the two, which indicates the applicability (or at least the relevance) of the Manning theory to the hydrated sulfonated polysulfones, and more generally, hydrated ionic polymers in the glassy state. The quantification of $b$ based on simulation outputs could enable a predictive estimate for the coion concentration in a membrane in contact with a salt solution, which would allow a forecast of a candidate polymer's capacity for coion exclusion.

\par Overall, our analyses indicate that the average inter-ion spacing, $b$, plays a central role in controlling the hydration and ion aggregation in an ionic polymer membrane. Here we have determined its value for two sulfonated polysulfones using computationally expensive MD simulations and revealed how the placement of the fixed ions on the polymer backbone subtly influences this parameter. Ideally, it is certainly desirable to construct a predictive theory of $b$ in terms of polymer characteristics including the ion content, the ion distribution along the backbone, the molecular weight distribution of the chains, the chain architecture, and the chain rigidity. Such a challenge can only be addressed with more systematic experimental and computational investigations.

\par We have estimated the degree of counterion condensation in dense ionic polymer membranes using the Manning theory (based on estimates for $b$ and $\varepsilon$) and have found a good agreement between the theoretical prediction and the fraction of counterions condensed to the fixed ions calculated directly from simulations. However, our analysis also highlights some of the challenges in applying the Manning theory as it was originally derived for a charged rod in a dilute salt solution.\cite{manning_limiting_1969} For a more thorough treatment, an approach based on the mean-field Poisson-Boltzmann theory may need to be considered.\cite{forsman_simple_2004, benyaakov_beyond_2009} A quantitative and broadly applicable theory for the coion concentration in membranes with high ion contents and/or in contact with concentrated salt solutions remains elusive.\cite{kamcev_effect_2017} Fundamental knowledge of the structure and thermodynamics of ionic aggregates in the concentrated (highly ionic) state is required to address this challenge.\cite{mceldrew_theory_2020}

\section*{Acknowledgement}
Acknowledgment is made to the Donors of the American Chemical Society Petroleum Research Fund (PRF \#56103-DNI6), for support of this research (C.W. and S.C.). This research is also supported by a 4-VA Collaborative Research Grant (S.C.) and the Virginia Tech College of Engineering (B.V. and J.J.L.). The authors acknowledge Advanced Research Computing at Virginia Tech (URL: http://www.arc.vt.edu) for providing computational resources and technical support that have contributed to the results reported within this paper.

%\bibliography{polysulfone}

%%%%% SUPPORTING INFORMATION %%%%%%%

%\clearpage
%\newpage
%\onecolumngrid
%\renewcommand{\floatpagefraction}{0.9}
%\renewcommand{\thefigure}{S\arabic{figure}}
%\setcounter{figure}{0}    
%\renewcommand{\thepage}{SI-\arabic{page}}
%\setcounter{page}{1}
\renewcommand{\theequation}{SI-\arabic{equation}}
\setcounter{equation}{0}   

\bigskip
\bigskip
\begin{center}
{\bf SUPPORTING INFORMATION}
\end{center}

\noindent \textbf{S1. Equilibrium Water Content}

\par The water content of the system is quantified as $\lambda$, which is defined as the number of water molecules per anion-cation pair. The value of $\lambda_\text{eq}$ in Table~1 of the main text for each polymer represents the value of $\lambda$ obtained when that polymer is at equilibrium with a liquid water environment at room temperature. We use the following equation to calculate $\lambda_\text{eq}$:
\begin{equation}
    \lambda_\text{eq}=\frac{f_\text{wu}*1000}{\text{IEC}*M_{\text{H}_2\text{O}}}~,
\end{equation}
where $f_\text{wu}$ is the equilibrium mass water uptake expressed as a fraction, IEC is the experimental ion exchange capacity in meq/g measured by titration, and $M_{\text{H}_2\text{O}}$ is the molar mass of water (18.015 g/mol). 

\noindent \textbf{S2. Film Casting}

\par Films are cast from the sulfonated polysulfones of varying ion contents starting with approximately 1 g of dry polymer dissolved in 20 mL of dimethylacetamide (DMAc) at room temperature. The solution is filtered through a 0.45 $\mu$m syringe filter and then sonicated for at least 10 minutes. A clean glass plate is placed on a level surface inside an airflow-controlled casting apparatus. This casting apparatus is similar to the one used in previous studies, but has the added feature of controlled airflow across the casting surface.\cite{daryaei2017-1, roy2017} The sonicated polymer solution is carefully poured onto the level casting plate. The air flow through the apparatus is set to 3 standard cubic feet per hour. Using a ceramic heating element (Tempco CRB30020) located above the casting surface, the temperature of the casting surface is increased to $\sim$45 \degree C. The temperature is increased again to $\sim$65 \degree C after 2 hours and further increased to $\sim$90 \degree C after an additional 2 hours. The film remains in the casting apparatus at 90 \degree C for around 12 hours. The casting plate with the solid film is then removed from the casting apparatus and placed in a vacuum oven. The film is held at 150 \degree C under vacuum for an additional 24 hours to remove any trace of DMAc. Finally, the casting plate with film is removed from the oven and allowed to cool down to room temperature. A bath of deionized (DI) water is used to aid in the delamination of the film from the glass plate. 

\noindent \textbf{S3. Ion Exchange}

\par The potassium form resulted from the original synthesis is exchanged to the sodium form by stirring the films in a 1M NaCl solution at room temperature overnight. The ion-exchanged films are then well rinsed and soaked in DI water for at least 48 hours, during which water is changed periodically to remove any excessive mobile ions. All films are stored in new DI water after casting. Water uptake and IEC of the films are measured using the same procedure that has been published previously.\cite{daryaei2017-1}

\noindent \textbf{S4. Measurement of Hydrated Polymer Density}

\par To measure the hydrated polymer density, the samples are again stored immersed in water before the measurement. A balance-based density determination kit is used. The samples are removed from water, with its surface quickly dried, and placed on the weighing pan to obtain the mass in air. The sample is then placed on the weighing basket in a bath of octadecene to obtain the mass in octadecene. The density of the hydrated polymer sample is computed using the following equation: 
\begin{equation}
    \rho_\text{p} = \frac{\rho_\text{o}*m_\text{a}-\rho_\text{a}*m_\text{o}}{m_\text{a} - m_\text{o}}~,
    \label{eq:wu}
\end{equation}
where $\rho_\text{o}$ is the density of octadecene at the measurement temperature, $m_\text{a}$ is the mass of the sample in air, $\rho_\text{a}$ is the density of air at the current elevation ($0.00101~\text{g/cm}^3$), and $m_\text{o}$ is the mass of the sample in octadecene. The octadecene density, $\rho_\text{o}$, is measured to be $0.7877~\text{g/cm}^3$ at ambient lab temperature using a 24 ml density bottle.

%%%%% END OF SUPPORTING INFORMATION %%%%%%%

\end{document}